\documentclass[aps,prc,twocolumn,showpacs,preprintnumbers,
               nofootinbib,float,superscriptaddress,longbibliography]{revtex4-1}
\usepackage{graphicx, fancybox}
\usepackage{amsmath,amssymb}
\usepackage{color}
\usepackage[dvipsnames]{xcolor}
\usepackage[colorlinks=true, pdfstartview=FitV, linkcolor=Orange, citecolor=Cyan, urlcolor=blue]{hyperref}
\usepackage{soul}
\usepackage{array}
\usepackage{url}
\usepackage[utf8]{inputenc}

\usepackage[normalem]{ulem}
\graphicspath{{figs/}}   

\begin{document}

\def\dNdeta{$dN^\mathrm{ch}/d\eta$}
\def\snn{\sqrt{s_\mathrm{NN}}}

\title{$\Lambda$ spin polarization in event-by-event relativistic heavy-ion collisions}

\author{Sahr Alzhrani}
\email{sahr.alzhrani@wayne.edu}
\affiliation{Department of Physics and Astronomy, Wayne State University, Detroit, Michigan, 48201, USA}
\affiliation{Department of Physics, Jazan University, Jazan, KSA}

\author{Sangwook Ryu}
\email{sangwook.ryu@wayne.edu}
\affiliation{Department of Physics and Astronomy, Wayne State University, Detroit, Michigan, 48201, USA}

\author{Chun Shen}
\email{chunshen@wayne.edu}
\affiliation{Department of Physics and Astronomy, Wayne State University, Detroit, Michigan, 48201, USA}
\affiliation{RIKEN BNL Research Center, Brookhaven National Laboratory, Upton, NY 11973, USA}

\begin{abstract}
We present a systematic study of $\Lambda$ hyperon's polarization observables using event-by-event (3+1)D relativistic hydrodynamics. The effects of initial hot spot size and QGP's specific shear viscosity on the polarization observables are quantified. We examine the effects of the two formulations of the thermal shear tensor on the polarization observables using the same hydrodynamic background. With event-by-event simulations, we make predictions for the Fourier coefficients of $\Lambda$'s longitudinal polarization $P^z$ with respect to the event planes of different orders of anisotropic flow. We propose new correlations among the Fourier coefficients of $P^z$ and charged hadron anisotropic flow coefficients to further test the mapping from fluid velocity gradients to hyperon's polarization. Finally, we present a system size scan with Au+Au, Ru+Ru, and O+O collisions at $\sqrt{s_\mathrm{NN}} = 200$\,GeV to study the system size dependence of polarization observables at the Relativistic Heavy-ion Collider.
\end{abstract}

{\maketitle}

\section{Introduction}

High energy nucleus-nucleus collisions at the Relativistic Heavy-Ion Collider (RHIC) and the Large Hadron Collider (LHC) create small droplets of Quark-Gluon Plasma (QGP), a hot and dense many-body system that carries the fundamental degrees of freedom of quarks and gluons. The QGP exhibits many intriguing emergent phenomena, such as nearly perfect fluidity and color opacity. Quantifying the QGP transport properties has been one of the primary goals of relativistic heavy-ion physics \cite{Gale:2013da, Shen:2020mgh}. Collisions with finite impact parameters carry large orbital angular momentum (OAM). This large OAM can induce local vorticity in the QGP fluid. Recently, the STAR Collaboration at RHIC discovered non-zero global polarization of $\Lambda$ hyperons, which indicated fluid vorticity of $\omega \approx (9 \pm 1) \times 10^{21} s^{-1}$ in semi-peripheral Au+Au collisions \cite{STAR:2017ckg}. This paradigm-shifting measurement together with the follow-up detailed analysis on differential global and longitudinal polarization observables \cite{STAR:2018gyt, STAR:2019erd} have opened a new venue to study the spin-related emergent properties of QGP at high energy.
Extensive theoretical and phenomenological investigations have been devoted to the effects of fluid vorticity on spin polarization \cite{Liang:2004ph, Becattini:2013fla, Becattini:2016gvu, Karpenko:2016jyx, Voloshin:2017kqp, Xie:2017upb, Karpenko:2021wdm, Huang:2020xyr, Becattini:2020ngo, Huang:2020dtn, Becattini:2020sww, Lisa:2021zkj, Serenone:2021zef, Becattini:2021lfq} as well as the related transport phenomenon involving spin \cite{Jiang:2016woz, Florkowski:2017ruc, Hattori:2019lfp, Liu:2019krs, Fukushima:2020ucl, Liu:2020ymh, Gao:2020vbh, Shi:2020htn, Li:2020eon, Singh:2020rht}. The (3+1)D hydrodynamics + hadronic transport hybrid models and multi-stage transport approaches can provide good descriptions of the global polarization for $\Lambda$ and $\bar{\Lambda}$ from the RHIC to LHC energies. 
However, theoretical calculations based on local thermal vorticity tensors showed the opposite oscillation pattern compared to the measured azimuthal distributions of polarization \cite{Becattini:2017gcx, Xia:2018tes, Florkowski:2019voj, Wu:2019eyi, Becattini:2019ntv}.

Recent works \cite{Hidaka:2017auj, Liu:2020dxg, Liu:2021uhn, Becattini:2021suc, Buzzegoli:2021wlg, Liu:2021nyg} proposed that the symmetric thermal shear tensor and gradients of $\mu_B/T$ can contribute to the spin polarization of $\Lambda$ and $\bar{\Lambda}$. There are two forms with the thermal shear tensor derived in the literature~\cite{Liu:2021uhn, Becattini:2021suc, Yi:2021ryh}.
The effects of thermal shear tensor on the longitudinal polarization's azimuthal dependence were studied and found to be substantial \cite{Fu:2021pok, Becattini:2021iol, Yi:2021unq}.
While it is still unclear which form should be suitable to use with the underlying hydrodynamic description, we will systematically study the effects of different shear-induced polarization terms on polarization observables using the same hydrodynamic background in this work.

Although there have been extensive works on hyperon's polarization observables, we find that studies about the sensitivity of polarization observables on the system's initial-state fluctuations and QGP viscosity are still lacking. Therefore, in this work, we perform (3+1)D event-by-event dynamical simulations of relativistic heavy-ion collisions at the top RHIC energy and systematically study how the polarization observables depend on the initial hot spot size and the QGP specific shear viscosity. We will compare these observables' sensitivity with those of other hadronic observables, such as anisotropic flow coefficients. With event-by-event simulations, we will further make model predictions for the azimuthal-dependent longitudinal polarization with respect to high-order event planes and new Pearson correlations between anisotropic flow and the Fourier coefficients of $P^z$ at fixed multiplicity.

This paper will be layout as follows. In Sec.~\ref{sec:model}, we will introduce our parametric 3D initial condition model based on Ref.~\cite{Shen:2020jwv, Ryu:2021lnx}. And we will summarize the two forms of symmetric thermal shear contributions to the fermion's polarization vector. We will express them using the same thermal shear tensor to highlight the difference. In Sec.~\ref{sec:results}, we study the sensitivity of hyperon's polarization observables on various medium parameters. We perform a system size scan with Au+Au, Ru+Ru, and O+O collisions at $\sqrt{s_\mathrm{NN}}$ = 200 GeV to study the system size dependence of polarization observables.
We will conclude with some closing remarks in Sec.~\ref{sec:conc}.

In this paper we use the conventions for the metric tensor $g^{\mu\nu} = \mathrm{diag}(1, -1, -1, -1)$ and the Levi-Civita symbol $\epsilon^{0123} = 1$.

\section{Model framework}\label{sec:model}

\subsection{3D event-by-event initial condition}

We employ the geometric-based 3D initial conditions developed in Ref.~\cite{Shen:2020jwv, Ryu:2021lnx} to carry out event-by-event simulations. Based on the Glauber geometry, the area density of energy and net longitudinal momentum at a given transverse position is given by,
\begin{eqnarray}
    \frac{d}{d^2 \textbf{x}_\perp} E(x, y) &=& [T_A(x, y) + T_B(x, y)] m_N \cosh(y_\mathrm{beam}) \nonumber \\
    &\equiv& M(x, y) \cosh(y_\mathrm{CM})\\
    \frac{d}{d^2 \textbf{x}_\perp} P_z(x, y) &=& [T_A(x, y) - T_B(x, y)] m_N \sinh(y_\mathrm{beam}) \nonumber \\
    &\equiv& M(x, y) \sinh(y_\mathrm{CM}).
\end{eqnarray}
Here $m_N$ is the nucleon mass, the beam rapidity is defined as $y_\mathrm{beam} \equiv \mathrm{arccosh}[\sqrt{s_\mathrm{NN}}/(2 m_N)]$, and $T_{A(B)}(x, y)$ is the participant thickness function in the transverse plane, composed by participant nucleons in the projectile(target) nucleus.
\begin{equation}
    T_{A(B)}(x, y) = \sum_{i \in A(B)} \frac{1}{2\pi w^2} \exp\left[- \frac{(x - x_i)^2 + (y - y_i)^2}{2 w^2}\right],
\end{equation}
where the $w$ parameter controls the hot spot size in the transverse plane. The summation of $i$ runs over all the participant nucleons inside the colliding nucleus. The lumpy profile of $T_{A(B)}(x, y)$ imprints its to the invariant mass and center-of-mass rapidity defined as follows,
\begin{eqnarray}
    M(x, y) &=& m_N \sqrt{T_A^2 + T_B^2 + 2 T_A T_B \cosh(2 y_\mathrm{beam})} \\
    y_\mathrm{CM}(x, y) &=& \mathrm{arctanh} \left[ \frac{T_A - T_B}{T_A + T_B} \tanh(y_\mathrm{beam}) \right].
\end{eqnarray}
Note that when $\cosh(2 y_\mathrm{beam}) \gg 1$, the local invariant mass scales with $\sqrt{T_A T_B}$ \cite{Shen:2020jwv}.

To study the polarization observables, it is essential to match the collision system's orbital angular momentum between the initial state and hydrodynamic fields event-by-event. This condition can be ensured by imposing the local energy and momentum conservation at every transverse position as follows,
\begin{eqnarray}
    && M(x, y) \cosh[y_\mathrm{CM}(x, y)] = \int \tau_0 d \eta_s [T^{\tau \tau}(x, y, \eta_s) \cosh(\eta_s) \nonumber \\
    && \qquad \qquad \qquad \qquad \qquad + \tau_0 T^{\tau \eta}(x, y, \eta_s) \sinh(\eta_s)] \label{eq:rule1} \\
    && M(x, y) \sinh[y_\mathrm{CM}(x, y)] = \int \tau_0 d \eta_s [T^{\tau \tau}(x, y, \eta_s) \sinh(\eta_s) \nonumber \\
    && \qquad \qquad \qquad \qquad \qquad + \tau_0 T^{\tau \eta}(x, y, \eta_s) \cosh(\eta_s)] \label{eq:rule2}.
\end{eqnarray}
Here $T^{\tau \tau}(x, y, \eta_s)$ and $T^{\tau \eta}(x, y, \eta_s)$ are components of the system's energy-momentum tensor on a constant proper time hyper-surface with $\tau = \tau_0$ where hydrodynamics starts.
Following Ref.~\cite{Ryu:2021lnx}, we assume the initial energy-momentum current has the following form,
\begin{eqnarray}
    T^{\tau\tau} (x, y, \eta_s) &=& e(x, y, \eta_s) \cosh(y_L) \label{eq:initial_Ttautau} \label{eq:Ttautau} \\
    T^{\tau \eta} (x, y, \eta_s) &=& \frac{1}{\tau_0} e(x, y, \eta_s) \sinh(y_L) \label{eq:Ttaueta}
    \label{eq:initial_Ttaueta}
\end{eqnarray}
with the initial longitudinal flow rapidity $y_L = f y_{\rm CM}$. The parameter $f$ controls the fraction of longitudinal momentum attributed to the initial longitudinal flow velocity. We ignore the transverse expansion and set transverse components $T^{\tau x} = T^{\tau y} = 0$ at $\tau = \tau_0$.
The longitudinal momentum fraction parameter $f$ allows us to vary the size of the initial longitudinal flow while keeping the net longitudinal momentum of the hydrodynamic fields fixed. In this work, we treat $f$ as a free parameter. It would be insightful to compare its optimal value with those from more sophisticated 3D initial conditions \cite{Karpenko:2016jyx, Shen:2017bsr, Fu:2021pok, Shen:2022oyg} in the future.  
Using Eqs.~\eqref{eq:rule1}-\eqref{eq:initial_Ttaueta}, we get
\begin{eqnarray}
    \!\!\!\!\!\!\!\!\!\!
    M(x, y) &=& \int \tau_0 d \eta_s e(x, y, \eta_s) \cosh(\eta_s - (y_\mathrm{CM} - y_L)) \\
    0 &=& \int \tau_0 d \eta_s e(x, y, \eta_s) \sinh(\eta_s - (y_\mathrm{CM} - y_L)).
\end{eqnarray}
To satisfy these two equations, we choose a symmetric rapidity profile parameterization w.r.t $y_\mathrm{CM} - y_L$ for the local energy density \cite{Hirano:2005xf},
\begin{eqnarray}
    && e (x, y, \eta_s; y_\mathrm{CM} - y_L) = \nonumber \\
    && \qquad \mathcal{N}_e(x, y) \exp\bigg[- \frac{(\vert \eta_s - (y_\mathrm{CM} - y_L) \vert  - \eta_0)^2}{2\sigma_\eta^2} \nonumber \\
    && \qquad \qquad \qquad \qquad \times \theta(\vert \eta_s - (y_\mathrm{CM} - y_L) \vert - \eta_0)\bigg].
    \label{eq:eprof}
\end{eqnarray}
Here the parameter $\eta_0$ determines the width of the plateau and the $\sigma_\eta$ controls how fast the energy density falls off at the edge of the plateau. The normalization factor $\mathcal{N}_e (x, y)$ is determined by the local invariant mass $M(x, y)$.
In a highly asymmetric situation $T_A(x, y) \gg T_B(x, y)$, the center-of-mass rapidity $y_\mathrm{CM}(x, y) \rightarrow y_\mathrm{beam}$. To make sure there is not too much energy density deposited beyond the beam rapidity, we set $\eta_0 = \mathrm{min}(\eta_0, y_\mathrm{beam} - (y_\mathrm{CM} - y_L))$. We include the same initial net baryon profiles as those in Refs.~\cite{Shen:2020jwv, Ryu:2021lnx}.

\begin{table*}[ht!]
    \centering
    \caption{The default choice of model parameters in the dynamic simulations for relativistic nuclear collisions at $\sqrt{s_\mathrm{NN}} = 200$\,GeV.}
    \begin{tabular}{l l c}
        \hline \hline
        Parameter \hspace{1cm} & Description \hspace{4cm} & Value \\ \hline
        $w$ [fm] & initial hot spot width & 0.4, 0.8, 1.2\\
        $\eta_0$ & space-time rapidity plateau size & 2.5 \\
        $\sigma_\eta$ & space-time rapidity fall off width & 0.5 \\
        $f$ & initial longitudinal flow fraction & 0.15 \\
        $\tau_0$ [fm/$c$] & hydrodynamics starting time & 1 \\
        $\eta T/(e+P)$ & specific shear viscosity & 0, 0.08, 0.16 \\
        $e_{\rm sw}$ [GeV/fm$^3$] & particlization energy density & 0.25, 0.5  \\ \hline \hline
    \end{tabular}
    \label{tab:modelParams}
\end{table*}

\subsection{(3+1)D dynamical evolution}

In this work, we use the open-source (3+1)D relativistic viscous hydrodynamic code package \textsc{music} \cite{Schenke:2010nt, Schenke:2011bn, Paquet:2015lta, Denicol:2018wdp, MUSIC} to simulate the dynamical evolution of the system’s energy, momentum, and net baryon density,
\begin{eqnarray}
    \partial_\mu T^{\mu\nu} = 0, \\
    \partial_\mu J_B^\mu = 0,
\end{eqnarray}
where the energy-momentum tensor is defined as
\begin{equation}
    T^{\mu\nu} = e u^\mu u^\nu - (P + \Pi) \Delta^{\mu\nu} + \pi^{\mu\nu}.
\end{equation}
The system's energy-momentum tensor is composed of the local energy density of the fluid cell $e$, the thermal pressure $P$, the fluid velocity $u^\mu$, and the shear stress tensor and bulk viscous pressure $\pi^{\mu\nu}$ and $\Pi$. The spatial projection tensor is defined as $\Delta^{\mu\nu} \equiv g^{\mu\nu} - u^\mu u^\nu$. Hydrodynamic equations of motion are solved with a lattice QCD based Equation of State (EoS) at finite baryon density \textsc{neos-bqs}, which imposes the strangeness neutrality condition and electric charge density $n_Q = 0.4 n_B$~\cite{Monnai:2019hkn}. The detailed equations of motion were explained in Ref.~\cite{Shen:2020jwv, Ryu:2021lnx}. In this work, we explore the shear viscous effects by running simulations with different values of $\eta T/(e+P)$ listed in Table~\ref{tab:modelParams}. This ratio reduces to $\eta/s$ at $\mu_B = 0$. We leave the influence of bulk viscosity for future work.

As the system evolves below the switching energy density $e_{\rm sw}$, individual fluid cells are converted to particles via the Cooper-Frye prescription~\cite{Cooper:1974mv, Huovinen:2012is, Shen:2014vra}. The produced hadrons further scatter with each other and decay in the hadronic phase, which is modeled by the hadronic transport model, \textsc{urqmd}~\cite{Bass:1998ca, Bleicher:1999xi}.

\subsection{Spin polarization of hyperons}

Recently, it is realized that the symmetric shear tensor contributes to the hyperon's polarization in addition to the anti-symmetric thermal vorticity tensor. Becattini {\it et. al.} \cite{Becattini:2021suc} derived a different thermal shear contribution compared to that from Ref.~\cite{Liu:2021uhn}. It is important to quantify the difference between these two formulations in phenomenological studies.

The $\Lambda$ hyperon's spin polarization vector as a function of its momentum $p^\mu$ can be computed as,
\begin{eqnarray}
    S^\mu(p^\alpha) = \frac{1}{4m} \frac{\int d \Sigma \cdot p\, n_0 (1 - n_0) \mathcal{A}^\mu }{\int d \Sigma \cdot p\, n_0},
    \label{eq:SpinPol}
\end{eqnarray}
where $m$ is the $\Lambda$'s mass and $n_0$ is the $\Lambda$'s momentum distribution at local thermal equilibrium. The $d\Sigma_\mu$ is the normal vector of the hyper-surface, on which we compute the hyperons' spin vectors. We use the same hyper-surface at the switching energy density $e_{\rm sw}$, on which the Cooper-Fyre particlization is performed. The axial vector $\mathcal{A}^\mu$ is composed by gradients of hydrodynamic fields.

On the one hand, Ref.~\cite{Becattini:2021suc} gives the following form for the axial vector,
\begin{eqnarray}
    \mathcal{A}_{\rm BBP}^\mu &=& - \varepsilon^{\mu\rho\sigma\tau} \left(\frac{1}{2} \omega_{\rho \sigma} p_\tau + \frac{1}{E} \hat{t}_\rho \xi_{\sigma \lambda}  p^\lambda p_\tau\right). \label{eq:Smu_SIP_BBP}
\end{eqnarray}
Here the first term represents the conventional contribution from the thermal vorticity tensor $\omega^{\mu\nu}$,
\begin{eqnarray}
    \omega^{\mu\nu} \equiv - \frac{1}{2} \left[\partial^\mu\left(\frac{u^\nu}{T}\right) - \partial^\nu\left(\frac{u^\mu}{T}\right) \right].
    \label{eq:thermalVorticityTensor}
\end{eqnarray}
The second term in Eq.~\eqref{eq:Smu_SIP_BBP} is the contribution from the symmetric thermal shear tensor. We denote this term as the Shear-Induced Polarization (SIP(BBP)). The global time-like vector $\hat{t}_\rho = (1, 0, 0, 0)$ and the thermal shear tensor $\xi^{\mu\nu}$ is defined as,
\begin{eqnarray}
    \xi^{\mu\nu} \equiv \frac{1}{2} \left[\partial^\mu\left(\frac{u^\nu}{T}\right) + \partial^\nu\left(\frac{u^\mu}{T}\right) \right].
    \label{eq:thermalShearTensor}
\end{eqnarray}
On the other hand, Ref.~\cite{Liu:2021uhn, Yi:2021ryh} proposed a different form for the shear-induced contribution, denoted as SIP(LY). To express it in terms of the thermal shear tensor $\xi^{\mu\nu}$ in Eq.~\eqref{eq:thermalShearTensor},
\begin{eqnarray}
    \mathcal{A}_{\rm LY}^\mu &=& - \varepsilon^{\mu\rho\sigma\tau} \left[  \frac{1}{2}\omega_{\rho \sigma} p_\tau  + \frac{1}{E} u_\rho \xi_{\sigma \lambda} p_\perp^\lambda p_\tau \right. \nonumber \\
    && \qquad \qquad  + \frac{b_i}{\beta E} u_\rho p^\perp_{\sigma} \partial^\perp_\tau (\beta \mu_B) \bigg].
    \label{eq:Smu_SIP_LY}
\end{eqnarray}
Note that there are two differences in the shear induced polarization in Eq.~\eqref{eq:Smu_SIP_LY} compared to that in Eq.~\eqref{eq:Smu_SIP_BBP}. First, Eq.~\eqref{eq:Smu_SIP_LY} uses the local flow velocity $u_\rho$ instead of a global time-like vector $\hat{t}_\rho$. Second, the shear-induced polarization in Eq.~\eqref{eq:Smu_SIP_LY} has an additional transverse projection operator acts on the momentum vector $p^\lambda$,
\begin{eqnarray}
    p_\perp^\lambda = p^\lambda - (u \cdot p) u^\lambda.
\end{eqnarray}
The flow velocity vector combined with the Levi-Civita tensor kills the temperature gradient terms from the thermal shear tensor and the transverse projection operator takes out the fluid acceleration terms from $\xi_{\sigma \lambda}$. 
The last term in Eq.~\eqref{eq:Smu_SIP_LY} represents the net baryon chemical potential induced polarization ($\mu_B$IP) \cite{Liu:2020dxg, Liu:2021uhn, Yi:2021ryh}.

Finally, the $\Lambda$'s polarization vector in the lab frame can be computed as
\begin{equation}
    P_{\rm lab}^\mu(p^\alpha) = \frac{1}{S} S^\mu(p^\alpha),
\end{equation}
where the spin $S = 1/2$ is for $\Lambda$ hyperons. Experimental measurements are often reported in the local rest frame of $\Lambda$,
\begin{equation}
    P^0 = 0 \quad \mbox{and} \quad P^i(p^\alpha) = P^i_\mathrm{lab} - \frac{\vec{p} \cdot \vec{P}_\mathrm{lab}}{p^0(p^0 + m)}p^i.
\end{equation}

In the following section, we will study the contribution from different axial vectors $\mathcal{A}^\mu$ in Eqs.~\eqref{eq:Smu_SIP_BBP} and \eqref{eq:Smu_SIP_LY} to the $\Lambda$'s polarization\footnote{Both axial vectors are numerically implemented in the open-source particle sampler \textsc{iss} to compute the hyperon's polarization observables~\cite{iSSpackage}.}. And we will explore how the polarization observables depend on the different sets of the dynamical model parameters listed in Table~\ref{tab:modelParams}.

\section{Results and discussion}\label{sec:results}

Our phenomenological study will first focus on Au+Au collisions at $\sqrt{s_{\rm NN}} = 200$\,GeV. After presenting the model calibration to the hadronic flow observables, we will dive into a systematic analysis of the hyperon's global and longitudinal polarization. With event-by-event simulations, we will make model predictions for the high-order oscillation patterns of longitudinal polarization. We will close this section by performing a system size scan for the $\Lambda$ polarization observables at the top RHIC energy.

\subsection{Particle production and flow observables}

Before studying the hyperon's polarization observables, we need to calibrate our model with the hadronic flow observables.

\begin{figure}[t!]
    \centering
    \includegraphics[width=0.95\linewidth]{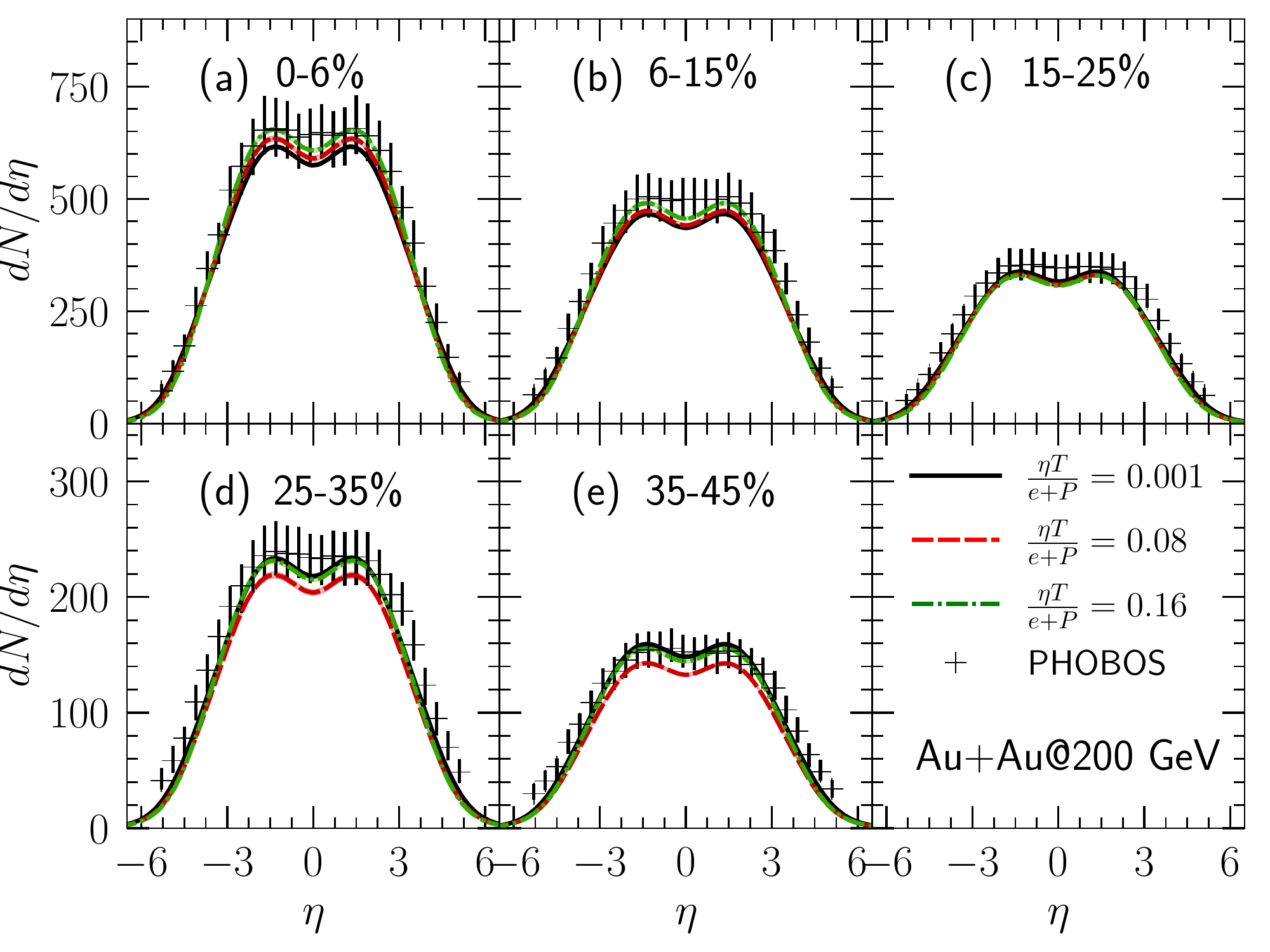}
    \includegraphics[width=0.95\linewidth]{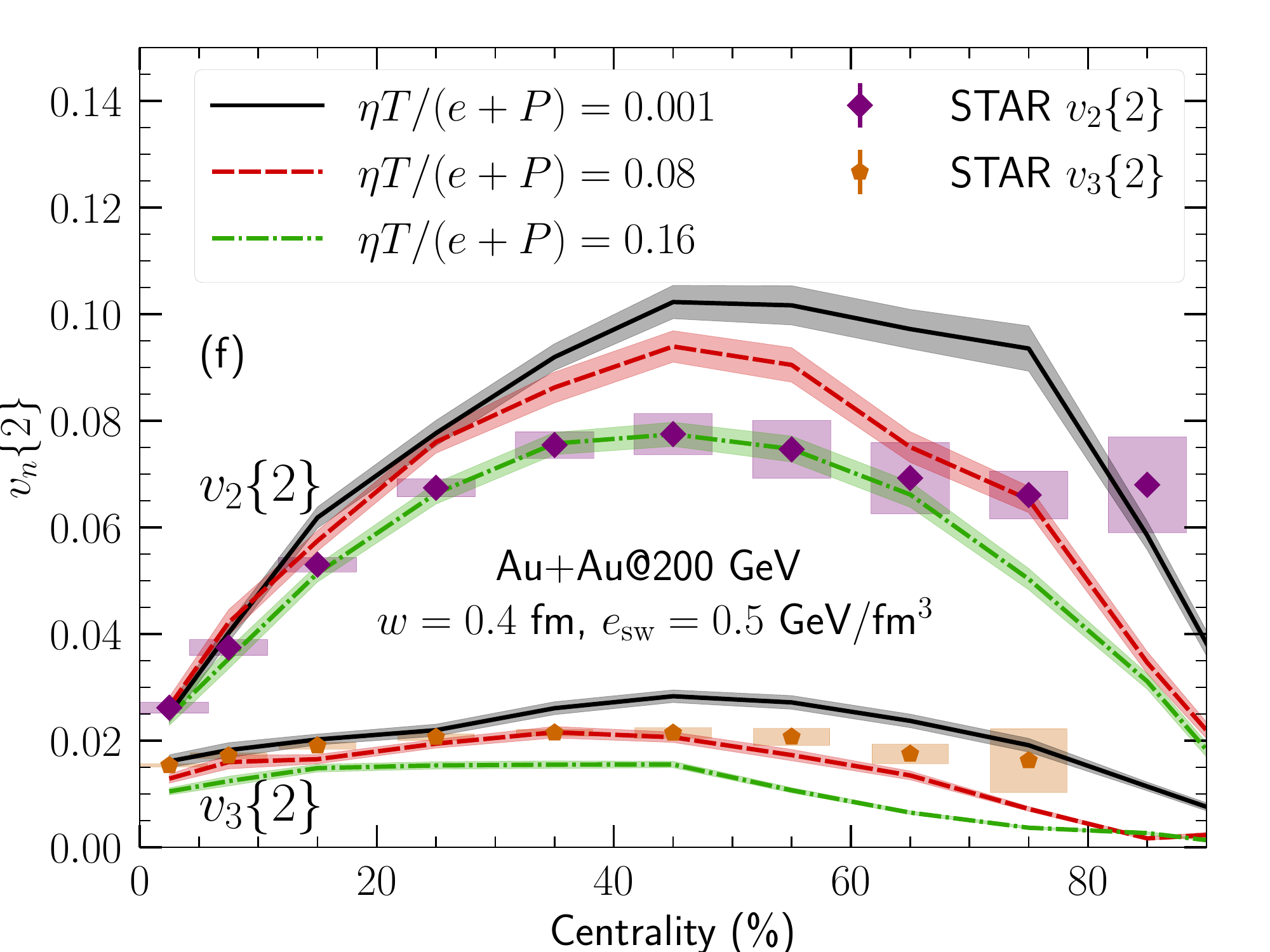}
    \caption{(Color Online) Panels (a)-(e): Charged hadron pseudo-rapidity distributions of Au+Au collisions at 200 GeV in different centrality bins compared to the PHOBOS measurements \cite{PHOBOS:2004vcu}. Panel (f): Charged hadron anisotropic flow coefficients compared with the STAR measurements \cite{STAR:2017idk}. Model simulations are performed with three values of specific shear viscosity in the hydrodynamic phase.}
    \label{fig:HadronRes}
\end{figure}

Figures~\ref{fig:HadronRes} show the results of our (3+1)D simulations for Au+Au collisions with different values of specific shear viscosity compared with experimental measurements. We adjust the space-time rapidity profile of the initial energy density to reproduce the PHOBOS charged hadron $dN^\mathrm{ch}/d\eta$ in 0-6\% centrality. The comparisons with the other semi-peripheral centrality bins show that our model can capture the centrality dependence of the particle production well. We note that the shear viscosity has small effects on the charged hadron pseudo-rapidity distributions. Meanwhile, the charged hadron anisotropic flow coefficients are suppressed more with larger specific shear viscosity. Comparing results from our initial condition with the hot spot size $w = 0.4$\,fm, we find that the $v_2\{2\}$ measurements favor a large value of specific shear viscosity $\eta T/(e + P) = 0.16$, while the $v_3\{2\}$ data prefers a smaller value $\eta T/(e + P) = 0.08$. Although the initial eccentricity follows the $\sqrt{T_A T_B}$ scaling as the popular \textsc{Trento} model \cite{Moreland:2014oya}, we do not include the hot spot normalization fluctuations which would increase $\epsilon_3$ relative to $\epsilon_2$ and would improve the overall description.

\begin{figure}[t!]
    \centering
    \includegraphics[width=0.95\linewidth]{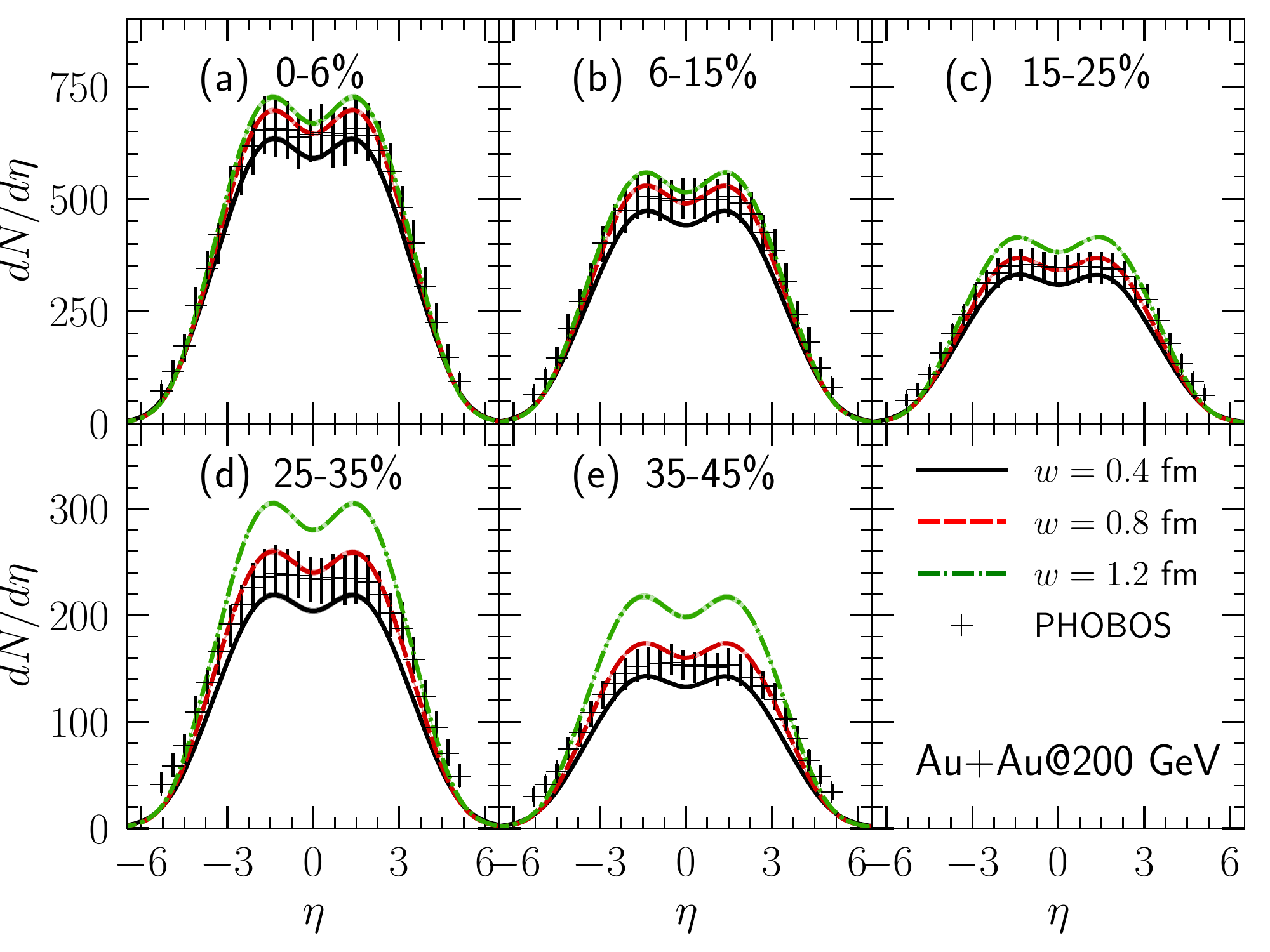}
    \includegraphics[width=0.95\linewidth]{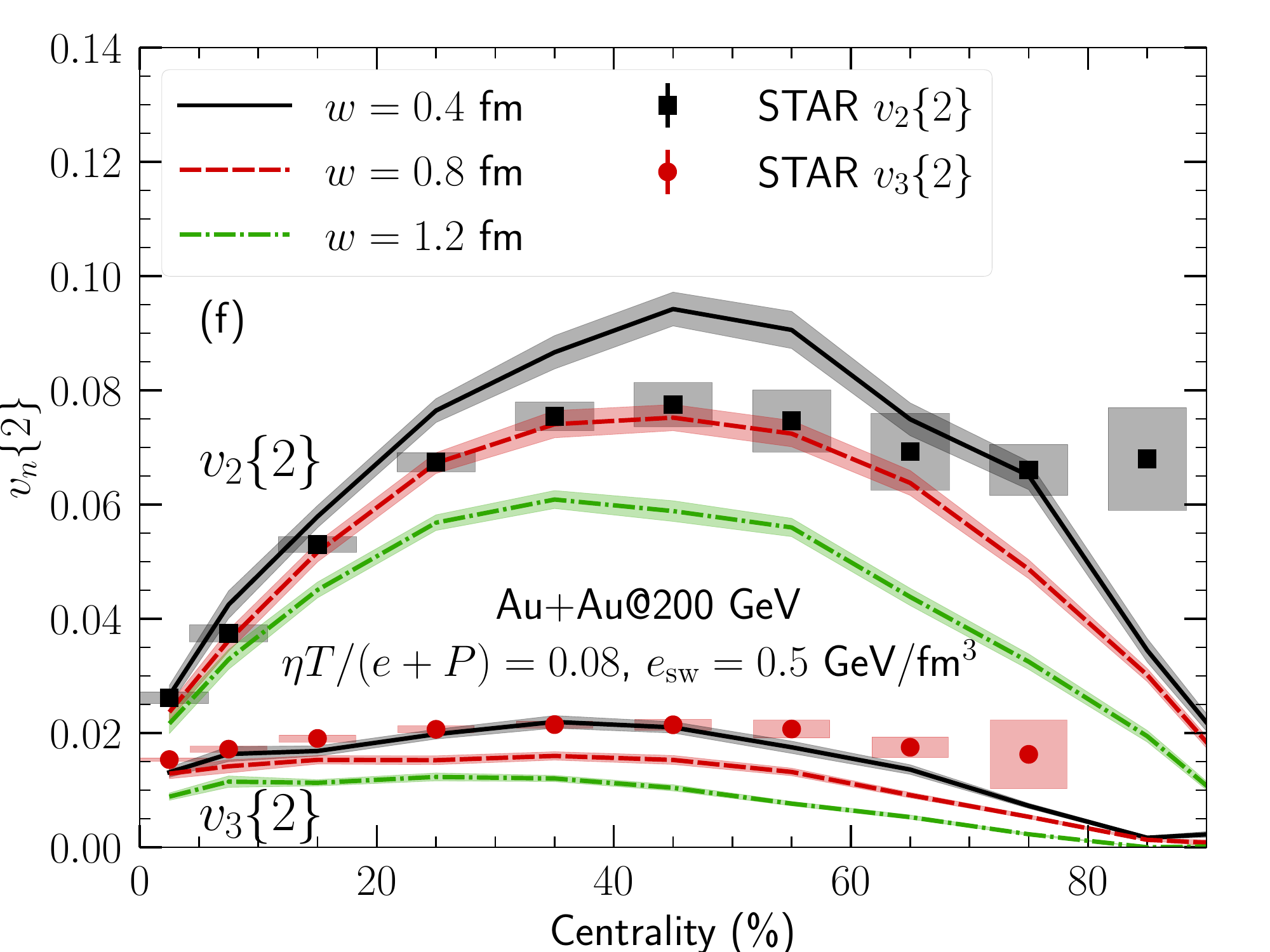}
    \caption{(Color Online) Similar comparisons as in Fig.~\ref{fig:HadronRes} but for model simulations with three initial hot spot sizes $w$.}
    \label{fig:HadronRes_wDep}
\end{figure}

Figures~\ref{fig:HadronRes_wDep} show how charged hadron $dN^\mathrm{ch}/d\eta$ and mid-rapidity anisotropic flow coefficients depend on the initial hot spot size in the simulations. We find that a large hot spot size $w = 1.2$ fm result in too much particle production near mid-rapidity in semi-peripheral collisions. The charged hadron $v_2\{2\}$ data prefers $w = 0.8$ fm, while the $v_3\{2\}$ data prefers $w = 0.4$ fm. Combining Figs.~\ref{fig:HadronRes} and \ref{fig:HadronRes_wDep}, we find the hadronic observables favor the initial hot spot size $w = 0.4 - 0.8$ fm and the specific shear viscosity $\eta T/(e+P) = 0.08 - 0.16$ with our model. The values of these parameters are consistent with previous works \cite{Shen:2020jwv, Giacalone:2021clp}.

\subsection{Global polarization of hyperons}

With the (3+1)D simulations calibrated to hadronic observables, we now start to study the hyperon's polarization observables.

\begin{figure}[ht!]
    \centering
    \includegraphics[width=0.9\linewidth]{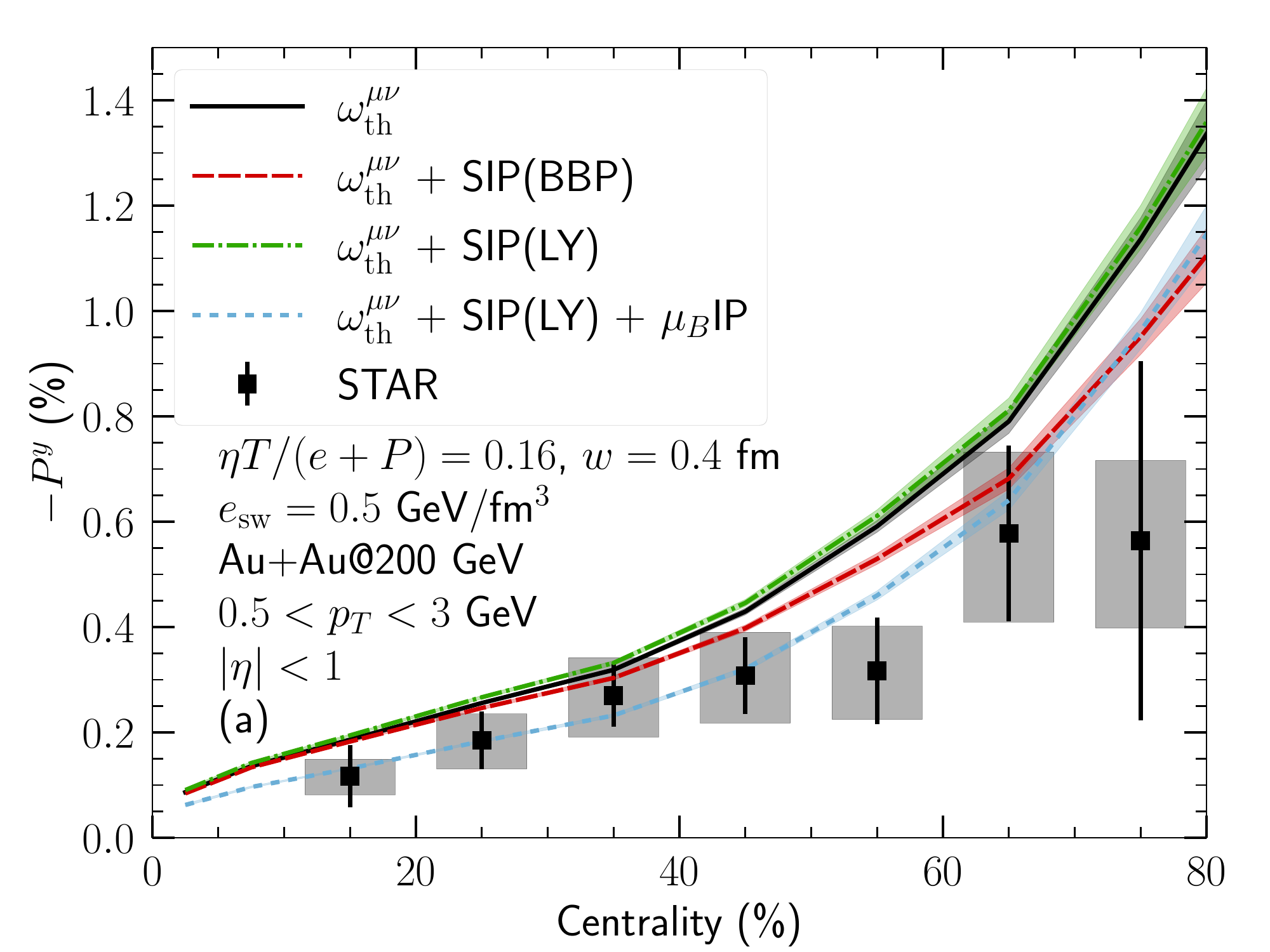}
    \includegraphics[width=0.9\linewidth]{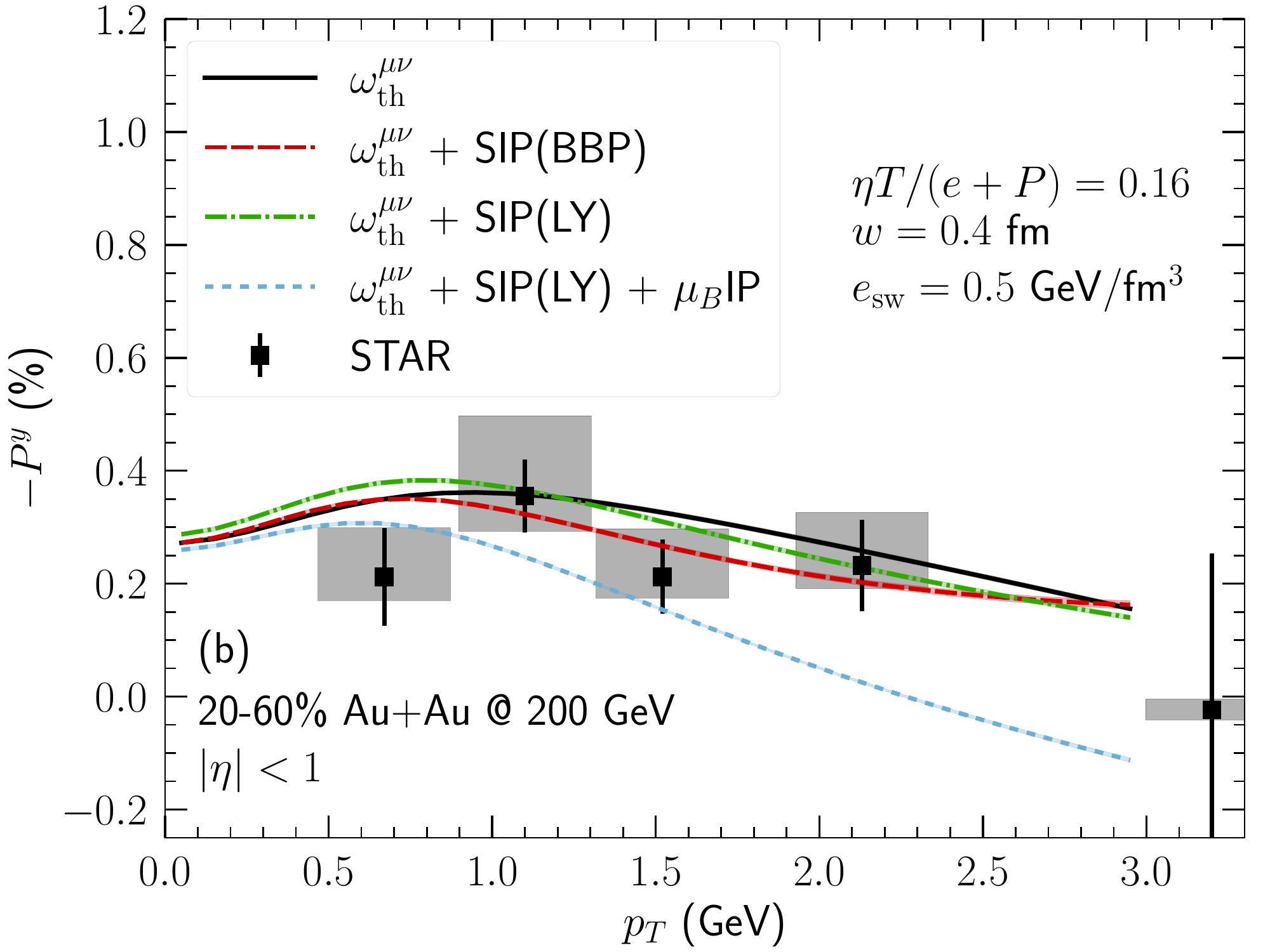}
    \includegraphics[width=0.9\linewidth]{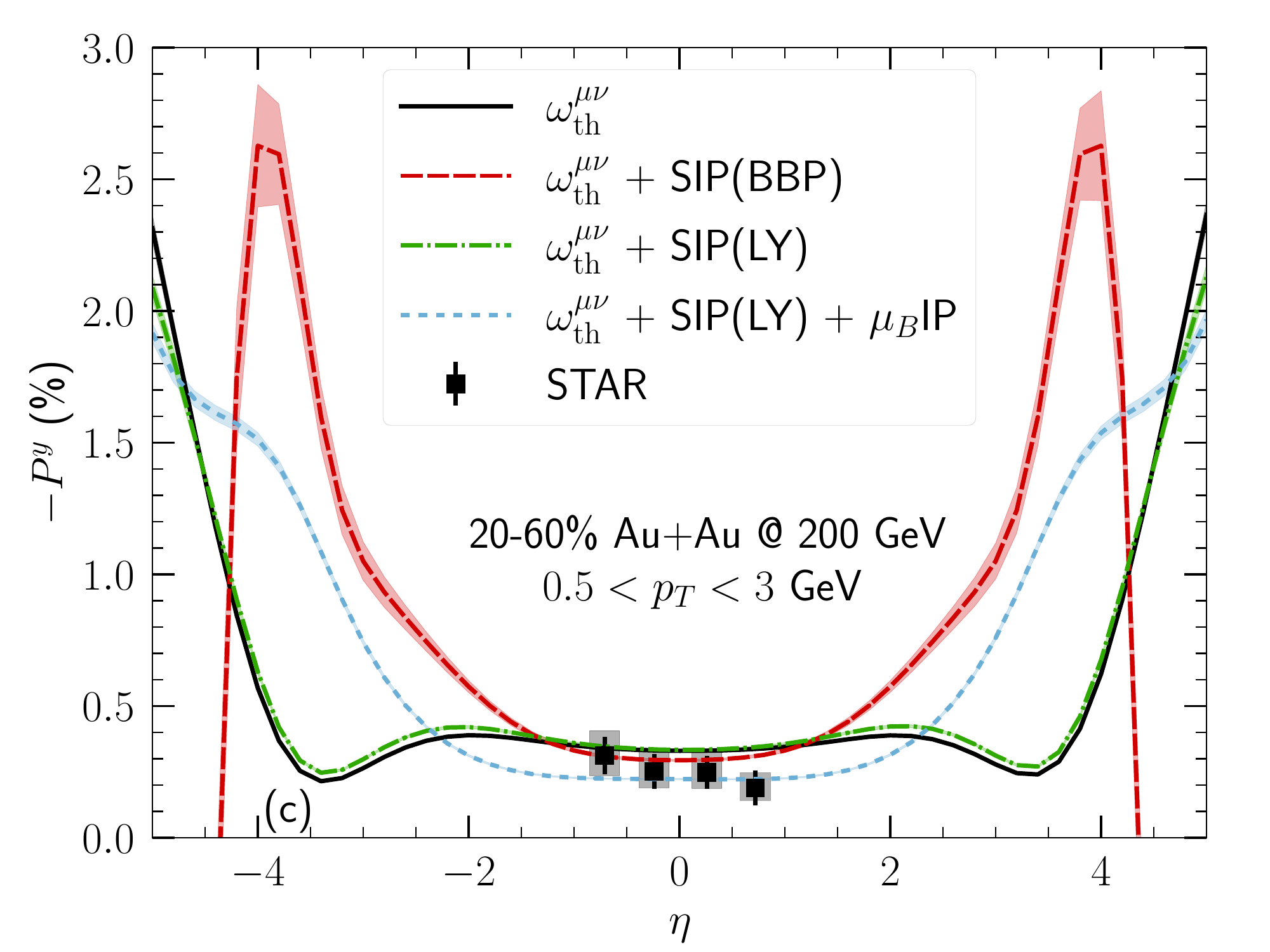}
    \caption{(Color Online) Panel (a): The hyperon's global polarization $P^y$ as a function of the collision centrality in Au+Au collisions at 200 GeV with four combinations of the axial-vector $\mathcal{A}^\mu$ in Eq.~\eqref{eq:SpinPol}. Panel (b): The $p_T$-differential $P^y$ with $\vert \eta \vert < 1$ in 20-60\% Au+Au collisions. Panel (c): The pseudo-rapidity distribution of $P^y$ in 20-60\% Au+Au collisions. Simulations are performed with the initial hot spot size $w = 0.4$ fm, a specific shear viscosity $\eta T/(e + P) = 0.16$ in the hydrodynamic phase, and a switching energy density $e_{\rm sw} = 0.5$\,GeV/fm$^3$. Comparisons are made with the STAR measurements \cite{STAR:2018gyt}. The STAR measurements are scaled by 0.877 because the latest hyperon decay parameter $\alpha_\Lambda$ from Ref.~\cite{ParticleDataGroup:2020ssz}.}
    \label{fig:PySIP}
\end{figure}

We first would like to quantify the four different combinations of the axial vectors' contributions to the $\Lambda$'s global polarization $P^y$. Because we define the $+x$ direction to be aligned with the impact parameter, the global angular momentum points to the $-y$ direction. Therefore, we plot $-P^y$ as $\Lambda$'s global polarization in all the following figures. Figure~\ref{fig:PySIP}a shows the $p_T$-integrated global polarization near mid-rapidity as a function of the collision centrality. We find $P^y$ is dominant by the thermal vorticity tensor, $\omega_\mathrm{th}^{\mu\nu}$ \cite{Ryu:2021lnx}. The symmetric thermal shear tensor proposed in Ref.~\cite{Becattini:2021suc} suppresses the magnitude of $P^y$ by 10\% in the semi-peripheral centrality bins. Meanwhile, the shear-induced polarization from Ref.~\cite{Liu:2021uhn} gives a negligible contribution to $P^y$. The tensor structure of $\mathcal{A}_{\rm LY}^\mu$ dictates a net-zero contribution to $P^y$ if $p_T$ is integrated from 0 to infinity. Figure~\ref{fig:PySIP}a shows that the contribution is negligible for integrating $p_T$ from 0.5 to 3 GeV. Lastly, the net baryon chemical potential induced polarization suppresses the $\Lambda$'s polarization, which agrees with the results using event-averaged initial conditions~\cite{Ryu:2021lnx}.

Although the two shear-induced polarization terms give small contributions to the $p_T$-integrated $P^y$, they have sizable effects on differential observables. Figure~\ref{fig:PySIP}b shows the $p_T$-differential $P^y(p_T)$ near mid-rapidity. The SIP(BBP) from Eq.~\eqref{eq:Smu_SIP_BBP} leads to a suppression of $P^y(p_T)$ for $p_T > 1$\,GeV. Meanwhile, the SIP(LY) from Eq.~\eqref{eq:Smu_SIP_LY} has a positive contribution for $p_T < 1$ GeV and negative contribution for $p_T > 1$ GeV. This sign-changing contribution ensures the $p_T$ integration of the SIP(LY) term in Eq.~\eqref{eq:Smu_SIP_LY} gives zero. The $\mu_B$-induced polarization suppresses the $\Lambda$'s $P^y$ to negative at high $p_T$. The current experimental uncertainty is still too large to discriminate between the thermal vorticity only case and the two shear-induced polarization. Figure~\ref{fig:PySIP}c shows the pseudo-rapidity dependence of the global polarization in 20-60\% Au+Au collisions. Although all four combinations of $\mathcal{A}^\mu$ give similar results near the mid-rapidity $\vert \eta \vert < 1.5$, the values of $P^y$ at the forward rapidity show a big difference. Measurements of $\Lambda$'s polarization at $\vert \eta \vert > 2$ would set strong constraints on the flow and net baryon chemical potential gradients in the hydrodynamic evolution.

\begin{figure}[t!]
    \centering
    \includegraphics[width=0.9\linewidth]{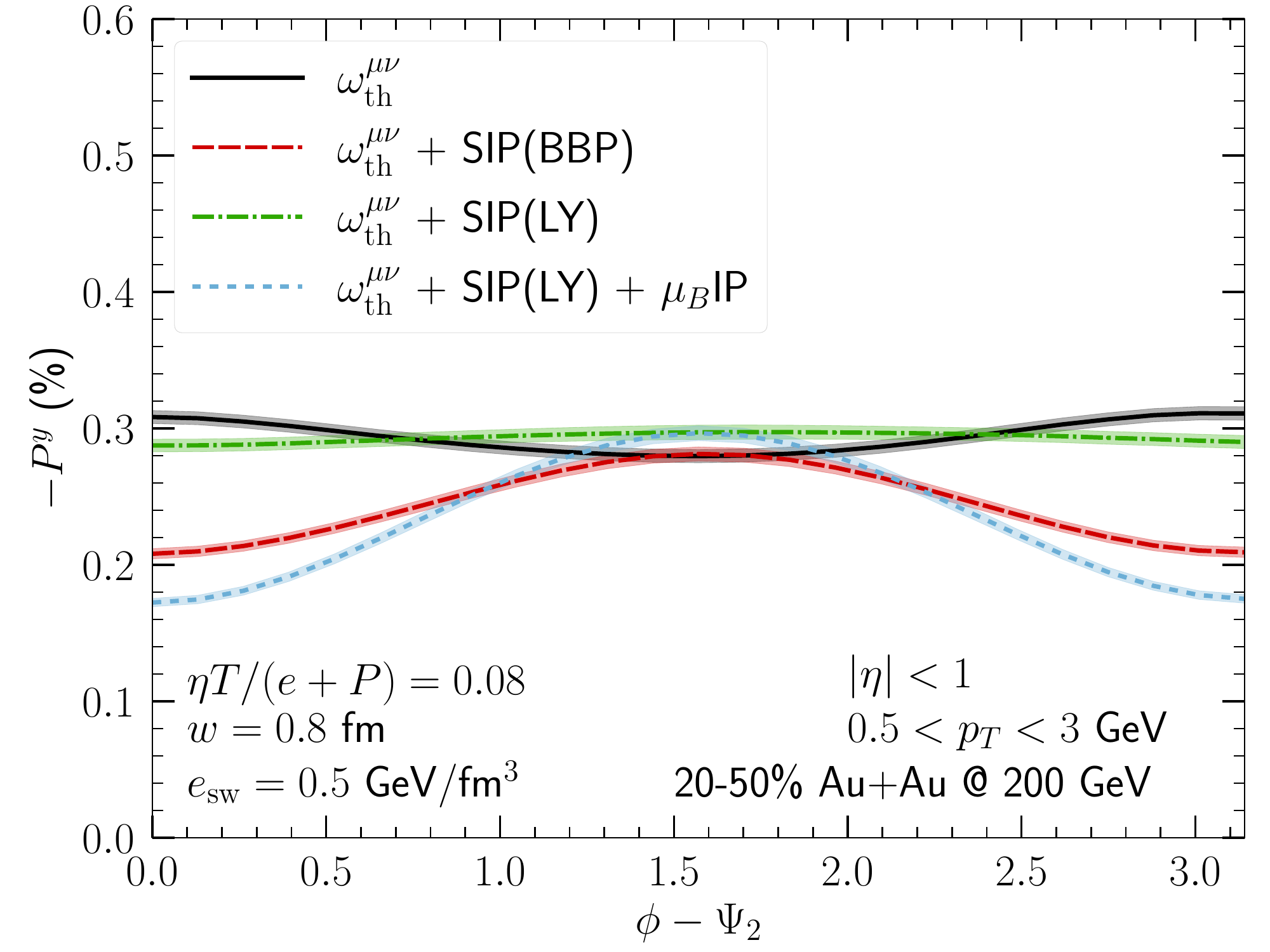}
    \caption{(Color Online) The azimuthal dependence of the global polarization $P^y$ with respect to the elliptic flow event plane for four combinations of the axial-vector $\mathcal{A}^\mu$.}
    \label{fig:PyPhiSIP}
\end{figure}

Figure~\ref{fig:PyPhiSIP} shows different polarization terms' contributions to the azimuthal dependence of the global polarization $P^y$ with respect to the elliptic flow event plane. The $n$-th order event plane angle $\Psi_n \equiv \mathrm{arg}(\mathcal{Q}_n)/n$ is defined by the complex flow vector $\mathcal{Q}_n \equiv Q_n e^{i n \Psi_n} = \sum_j e^{i n \phi_j}$, where $j$ runs over the azimuthal angles of all charged hadrons within the desired kinematic range. In our hybrid simulations, we sample multiple hadronic events from the same hydrodynamic hyper-surface to gain enough resolution of the anisotropic flow event-plane $\Psi_n$ for every hydrodynamic event \cite{McDonald:2016vlt}.
With our 3D initial conditions, the thermal vorticity generate a small but positive $\cos(2(\phi - \Psi_2))$ modulation of $P^y$. This result is different from simulations using other types of initial conditions based on transport models, such as UrQMD or AMPT \cite{Karpenko:2016jyx, Fu:2021pok}. The shear induce polarization gives the opposite $\cos(2\phi)$ modulation to that from the thermal vorticity. The SIP(BBP) from Ref.~\cite{Becattini:2021suc} gives a larger contribution compared to that from the SIP(LY) from Ref.~\cite{Liu:2021uhn}. Finally, we find that the chemical potential induced polarization term also gives a substantial contribution to the azimuthal dependence of $P^y$. Its contribution is related to the shape of the initial net baryon profile in the initial-state model. A recent work \cite{Fu:2022myl} investigated the $\mu_B$IP as a function of collision energy using the AMPT initial conditions.

\begin{figure}[t!]
    \centering
    \includegraphics[width=0.9\linewidth]{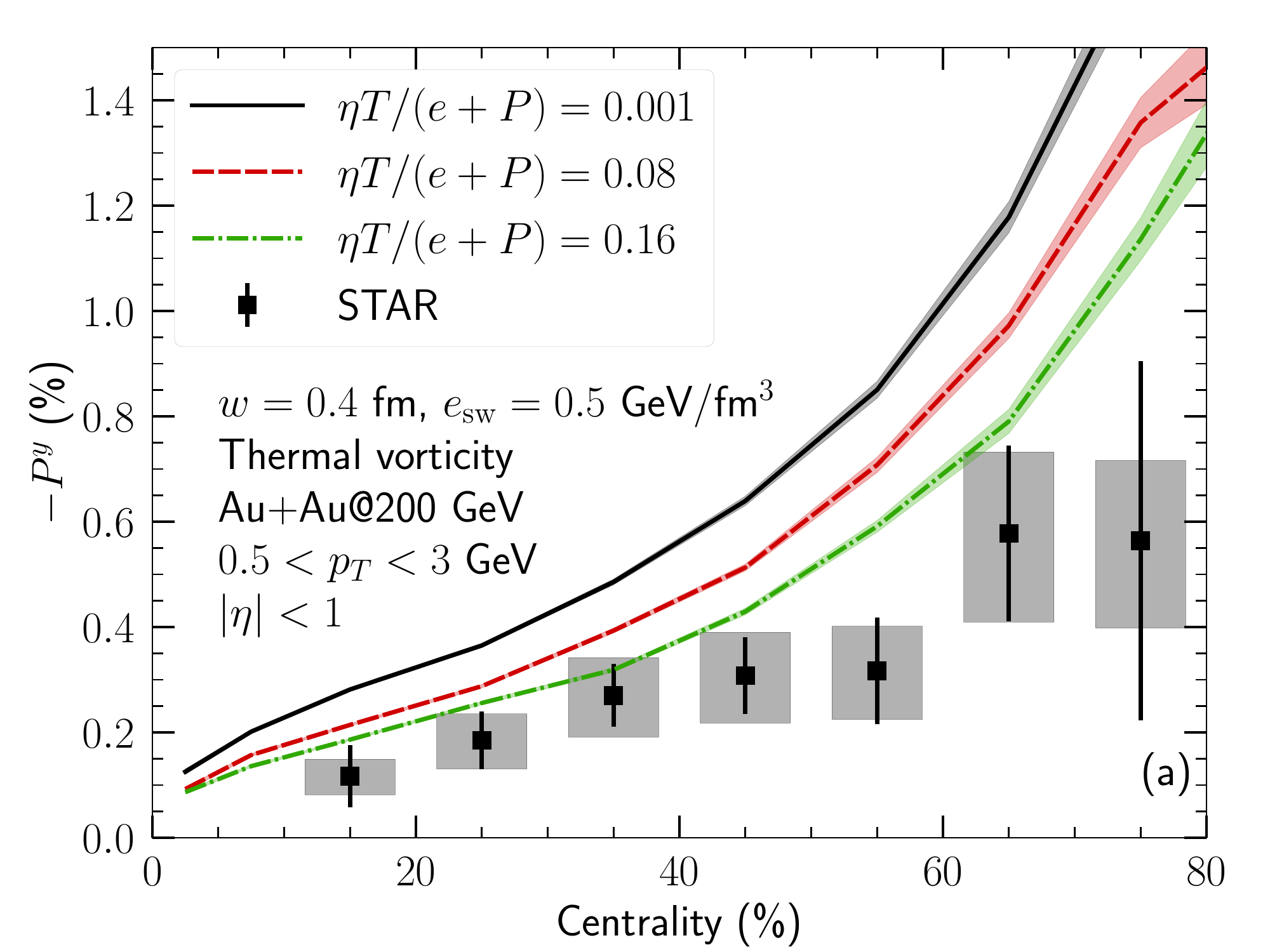}
    \includegraphics[width=0.9\linewidth]{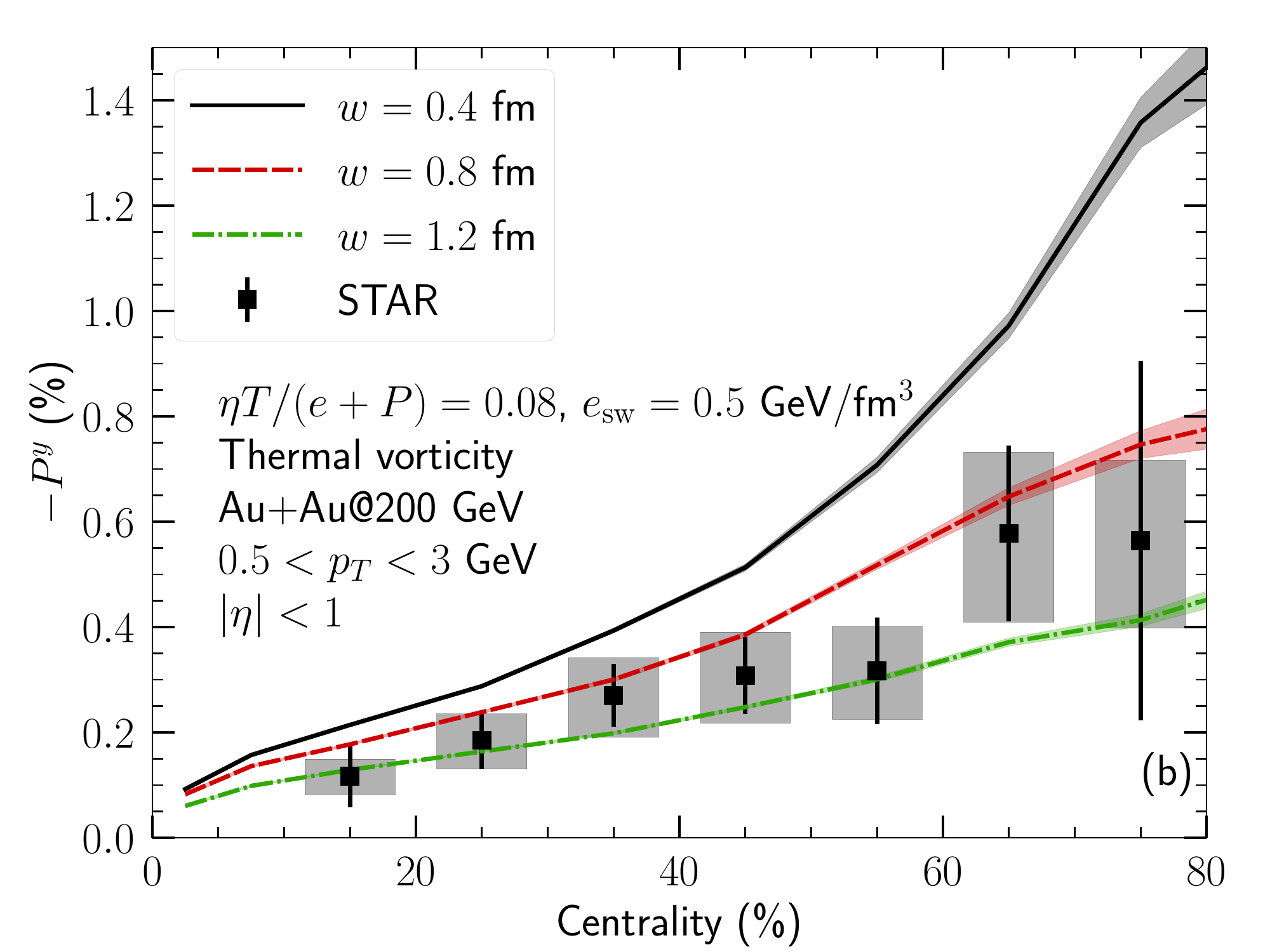}
    \includegraphics[width=0.9\linewidth]{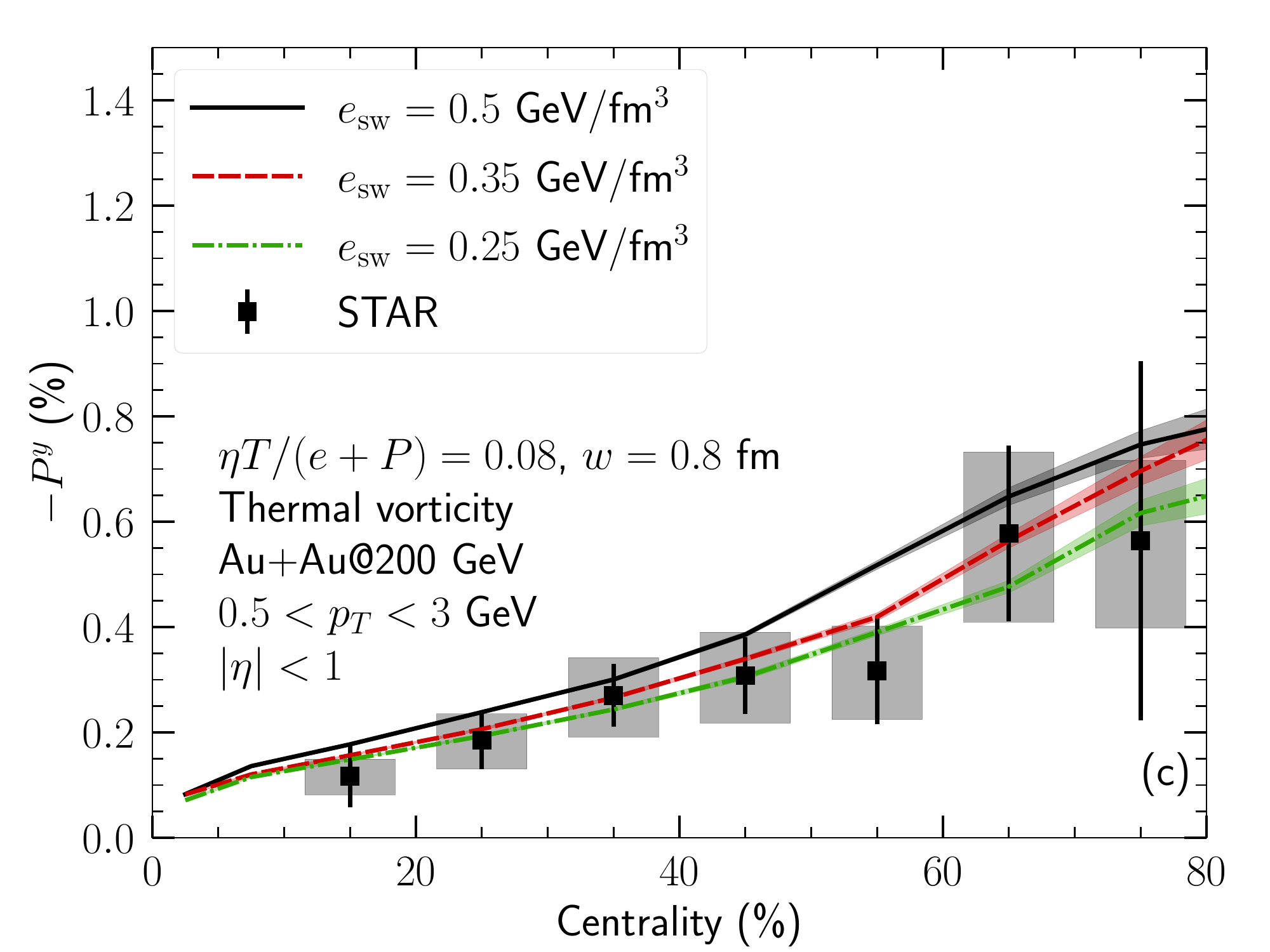}
    \caption{(Color Online) The sensitivity of hyperon's global polarization on the QGP's specific shear viscosity (a), initial hot spot size (b), and switching energy density (c). The STAR measurements \cite{STAR:2018gyt} are scaled by 0.877 because the latest hyperon decay parameter $\alpha_\Lambda$ from Ref.~\cite{ParticleDataGroup:2020ssz}.}
    \label{fig:PyParamDep}
\end{figure}

Figure~\ref{fig:PyParamDep}a shows how the $\Lambda$'s polarization depends on the specific shear viscosity used in the hydrodynamic phase. Similar to the anisotropic flow coefficients, a large specific shear viscosity leads to a significant suppression the global polarization. This result is expected because shear viscosity smears out the flow velocity gradients and the simulations will end up with smaller vorticity on the particlization surface. Comparing the relative magnitudes of suppression in polarization and anisotropic flow in Fig.~\ref{fig:HadronRes}, we find they are comparable.
Figure~\ref{fig:PyParamDep}b shows a substantial sensitivity of the $\Lambda$'s global polarization on the initial hot spot size. A smaller hot spot size leads to larger spatial gradients at the early time, which build up the stronger hydrodynamic flow. Therefore, a small $w$ results in larger thermal vorticity at the particlization surface in the simulations and enhance the magnitudes of the $\Lambda$'s global polarization.
Figure~\ref{fig:PyParamDep}c further explores how the global polarization depends on the switching energy density. A lower switching energy density allows the fireball to evolve longer. The flow velocity gradients reduce with $e_\mathrm{sw}$. Our results are  in qualitative agreement with the recent work \cite{Sun:2021nsg}.

The parameter dependence studies presented in Figures~\ref{fig:PyParamDep} demonstrate that the global polarization observables have a strong sensitivity to the initial-state fluctuations and QGP's specific shear viscosity. Combining the knowledge from hadronic observable comparisons in Figs.~\ref{fig:HadronRes} and \ref{fig:HadronRes_wDep}, we can draw tighter constraints on modeling the dynamical evolution of relativistic heavy-ion collisions. 

\subsection{Azimuthal-dependent longitudinal polarization}

Now, we transit our focus to longitudinal polarization $P^z$, which is sensitive to the flow velocity distribution in the transverse plane \cite{Voloshin:2017kqp}.

\begin{figure}[t!]
    \centering
    \includegraphics[width=0.9\linewidth]{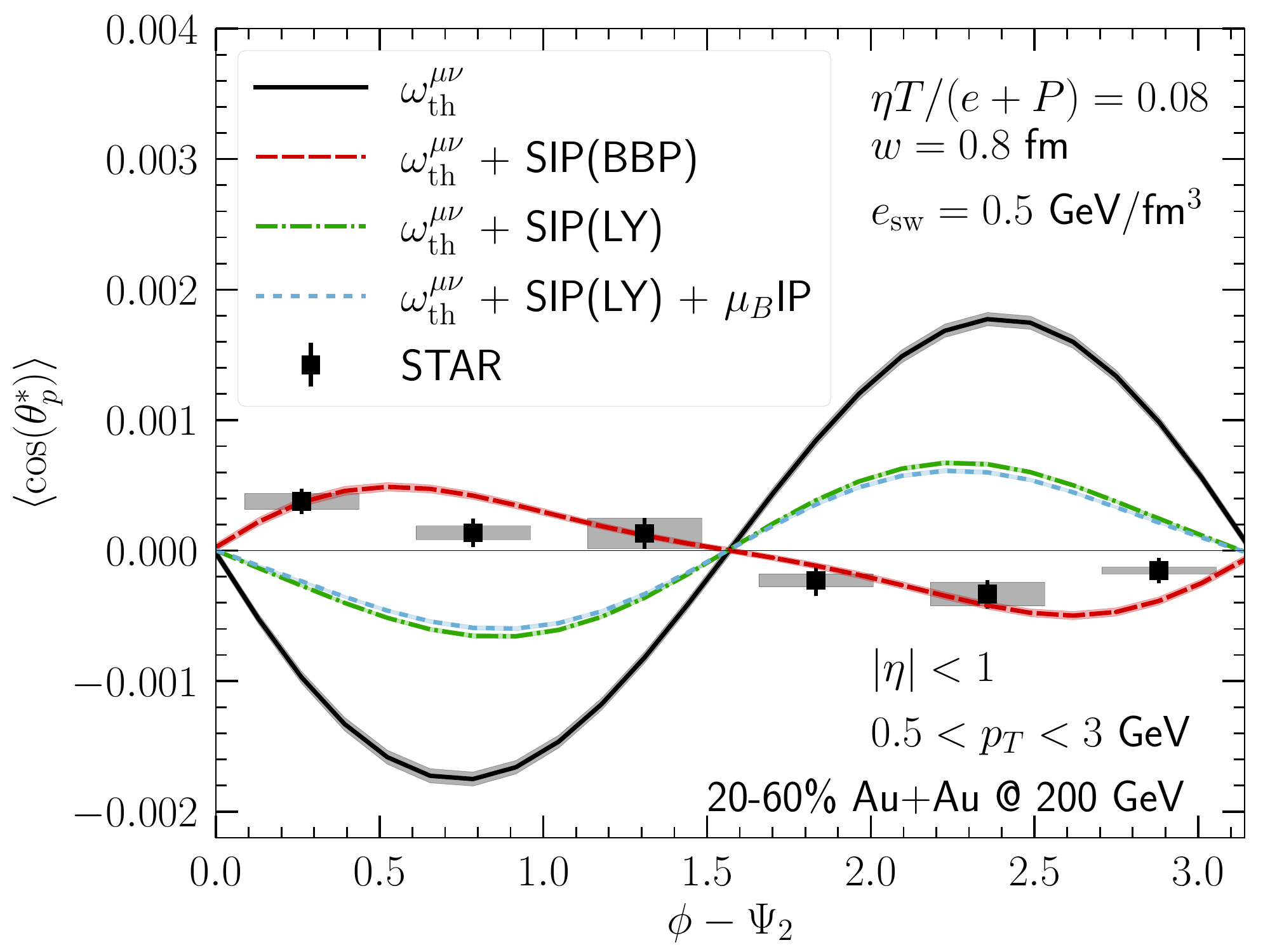}
    \caption{(Color Online) The averaged cosine of the daughter proton's polar angle in the $\Lambda$'s rest frame computed from $\Lambda$'s longitudinal polarization with four combinations of the axial-vector $\mathcal{A}^\mu$ in 20-60\% Au+Au collisions. Model calculations are compared with the STAR measurements \cite{STAR:2019erd}.}
    \label{fig:PzSIP}
\end{figure}

Figure~\ref{fig:PzSIP} shows the azimuthal dependence of the averaged cosine of the daughter proton's polar angle $\theta_p^*$ in the $\Lambda$'s rest frame with respect to the elliptic flow event plane. We compute this observable from the $\Lambda$'s longitudinal polarization $P^z$,
\begin{equation}
    \langle \cos(\theta_p^*) \rangle(\phi) = \langle \cos(\theta_p^*)^2 \rangle \alpha_\Lambda  P^z(\phi),
    \label{eq:costhetap}
\end{equation}
where $\langle \cos(\theta_p^*)^2 \rangle = 1/3$ \cite{STAR:2019erd} and $\alpha_\Lambda = 0.732$ \cite{ParticleDataGroup:2020ssz}.
The azimuthal-dependent longitudinal polarization $P^z(\phi)$ are computed using the four combinations of the axial-vector $\mathcal{A}^\mu$. Similar to previous works, the thermal vorticity alone gives the opposite sign of the $\phi$-dependence compared to the STAR measurements. The scale of the oscillation is about 5 times bigger than that in the data. Adding the shear-induced polarization from Ref.~\cite{Becattini:2021suc} flips the sign of the longitudinal polarization. While the sign of the SIP correction agrees with the results shown in Ref.~\cite{Becattini:2021iol}, the magnitude of the correction is bigger in our calculations. We believe the difference lies in the different types of initial conditions used in the simulations. The shear-induced polarization from Ref.~\cite{Liu:2021uhn} gives a smaller contribution compared to that from the SIP(BBP) term. Our results with the SIP(LY) are in quantitative agreement with those shown in Ref.~\cite{Fu:2021pok, Yi:2021unq}. The difference between the two SIP terms can be understood as the flow velocity vector $u_\rho$ combined with the Levi-Civita tensor killing the contributions from the temperature gradients in the thermal shear tensor. And the transverse projection operator on $p^\perp_\lambda$ in Eq.~\eqref{eq:Smu_SIP_LY} takes out the fluid acceleration contributions. These two contributions are substantial enough to change the sign of the longitudinal polarization within our model. Lastly, the net baryon chemical potential gradients give small contributions to $\Lambda$'s longitudinal polarization.  

After quantifying the individual term's contribution from the axial vector to $\Lambda$'s longitudinal polarization, we compare our model calculations with the STAR data as a function of the collision centrality \cite{STAR:2019erd}. We expand the longitudinal polarization $P^z(\phi)$ into a Fourier series as follows,
\begin{equation}
    P^z(\phi) = P_0^z + 2 \sum_{n = 1}^{\infty} P_n^z \cos(n(\phi - \Psi_n^{P^z})).
\end{equation}
Here the $n$-th order Fourier coefficient and its associated phase can be combined as a complex vector,
\begin{equation}
    \mathcal{P}_n^z \equiv P_n^z e^{i n \Psi_n^{P^z}} \equiv \int_0^{2\pi} \frac{d \phi}{2\pi} P^z(\phi) e^{i n \phi}.
\end{equation}
In heavy-ion experiments, one measures the magnitude of the $P^z$ oscillation with respect to the event plane angle defined by the charged hadron anisotropic flow vector,
\begin{eqnarray}
    && \!\!\!\!\!\!\!\!\!\!\!\!\! \langle P^z \sin(n(\phi - \Psi_n)) \rangle  \nonumber \\
    &=& \frac{1}{N_\mathrm{ev}} \sum_{i=1}^{N_\mathrm{ev}} \frac{1}{2\pi} \int^{2\pi}_0 d\phi P_i^z(\phi) \sin(n(\phi - \Psi_{i,n})) \nonumber \\
    &=& \left\langle \mathrm{Im} \left\{\mathcal{P}_n^z \frac{\mathcal{Q}_n^*}{\vert \mathcal{Q}_n \vert} \right\}  \right\rangle_\mathrm{ev}.
\end{eqnarray}
Here the $\mathcal{Q}_n$ is the complex anisotropic flow vector of charged hadrons and the operator $\mathrm{Im}\{\cdots\}$ takes the imaginary part of the enclosed expression.
The event average goes over all hydrodynamic events within a given centrality bin. In the low event-plane resolution limit \cite{Luzum:2012da},
\begin{equation}
    \langle P^z \sin(n(\phi - \Psi_n)) \rangle \simeq p_n^z\{\mathrm{SP}\} \equiv \frac{\left\langle \mathrm{Im} \left\{\mathcal{P}_n^z \mathcal{Q}_{n, A}^* \right\}  \right\rangle_\mathrm{ev}}{\sqrt{\langle \mathrm{Re}\{ \mathcal{Q}_{n, A} \mathcal{Q}_{n, B}^*\}\rangle_\mathrm{ev}}}.
\end{equation}
Here $\mathcal{Q}_{n, A}$ and $\mathcal{Q}_{n, B}$ are the anisotropic flow vectors from two sub-events. In the following analysis, we choose sub-event $A$ with charged hadrons whose $p_T \in [0.2, 3]$\,GeV and $\eta \in [-1, -0.1]$ and sub-event $B$ with charged hadrons having $p_T \in [0.2, 3]$\,GeV and $\eta \in [0.1, 1]$. 

\begin{figure}[t!]
    \centering
    \includegraphics[width=0.9\linewidth]{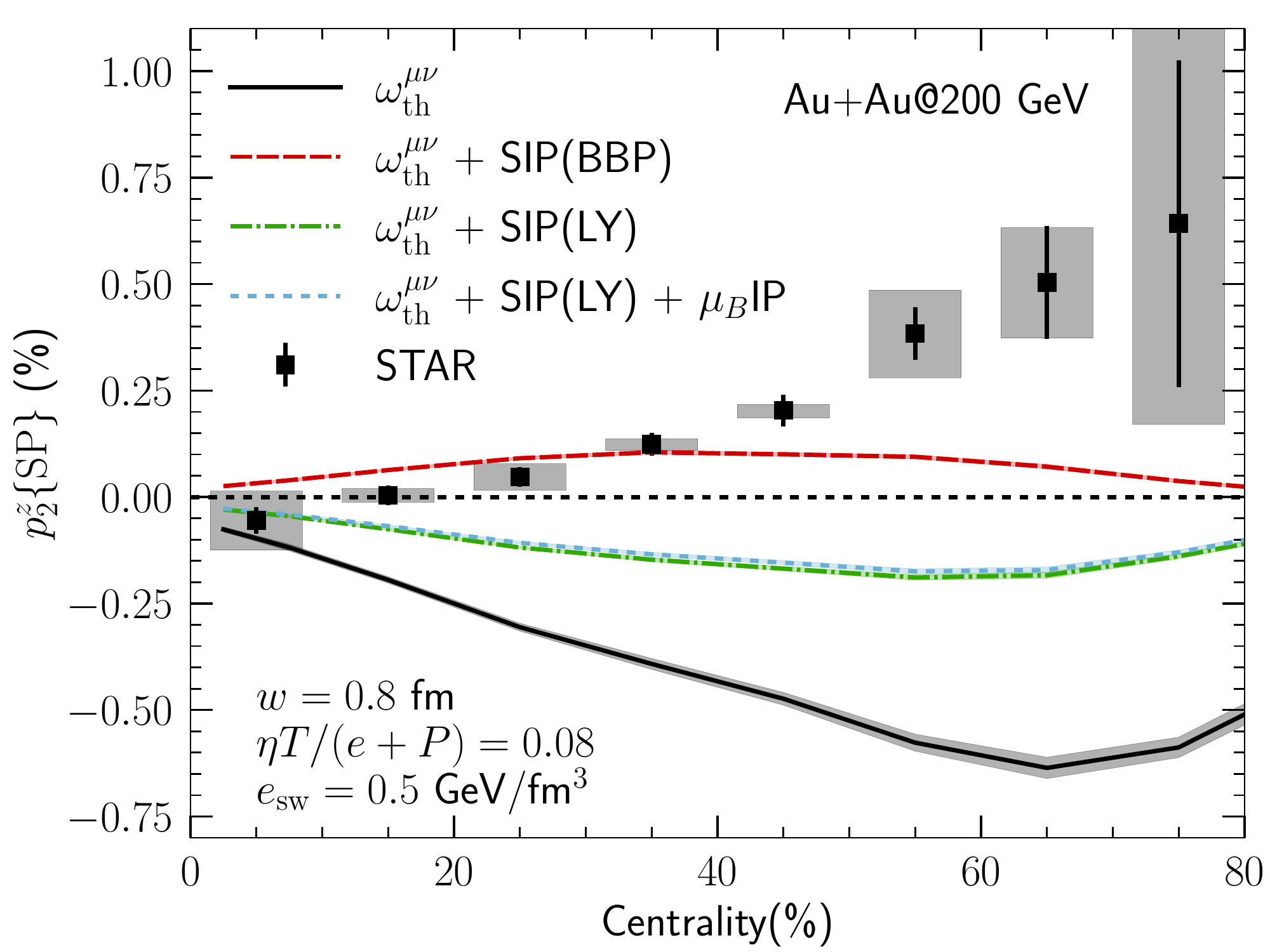}
    \caption{(Color Online) The centrality dependence of the second-order Fourier coefficients of the azimuthal dependent $P^z$ with respect to the elliptic flow event plane angle $\Psi_2$ for four combinations of the axial-vector $\mathcal{A}^\mu$. Results are compared with the STAR data \cite{STAR:2019erd}.}
    \label{fig:PzCenSIPDep}
\end{figure}

Figure~\ref{fig:PzCenSIPDep} shows that the results from thermal vorticity alone and those with adding the shear-induced polarization from Ref.~\cite{Liu:2021uhn} give negative values for the second-order Fourier coefficients of $P^z(\phi)$ with respect to the elliptic flow event plane. The thermal shear tensor with the SIP(BBP) from Ref.~\cite{Becattini:2021suc} gives positive results for $p_2^z\{\mathrm{SP}\}$. Comparing these results with the STAR measurements, we find reasonable agreements from central up to 40\% centrality. The magnitude of $p_2^z\{\mathrm{SP}\}$ in our calculation starts to decrease in peripheral centrality bins, while the measurement values keep increasing.

\begin{figure}[t!]
    \centering
    \includegraphics[width=0.9\linewidth]{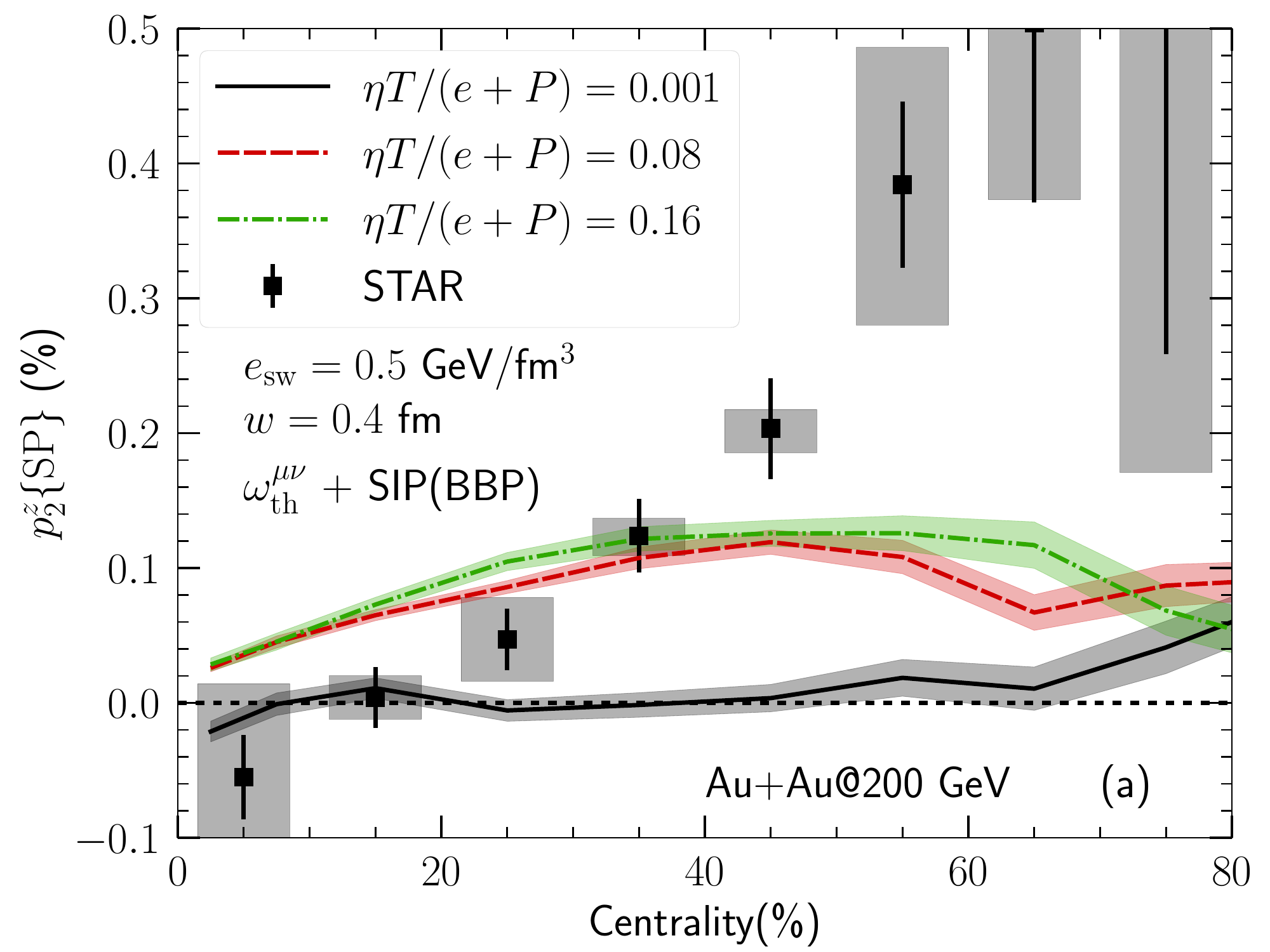}
    \includegraphics[width=0.9\linewidth]{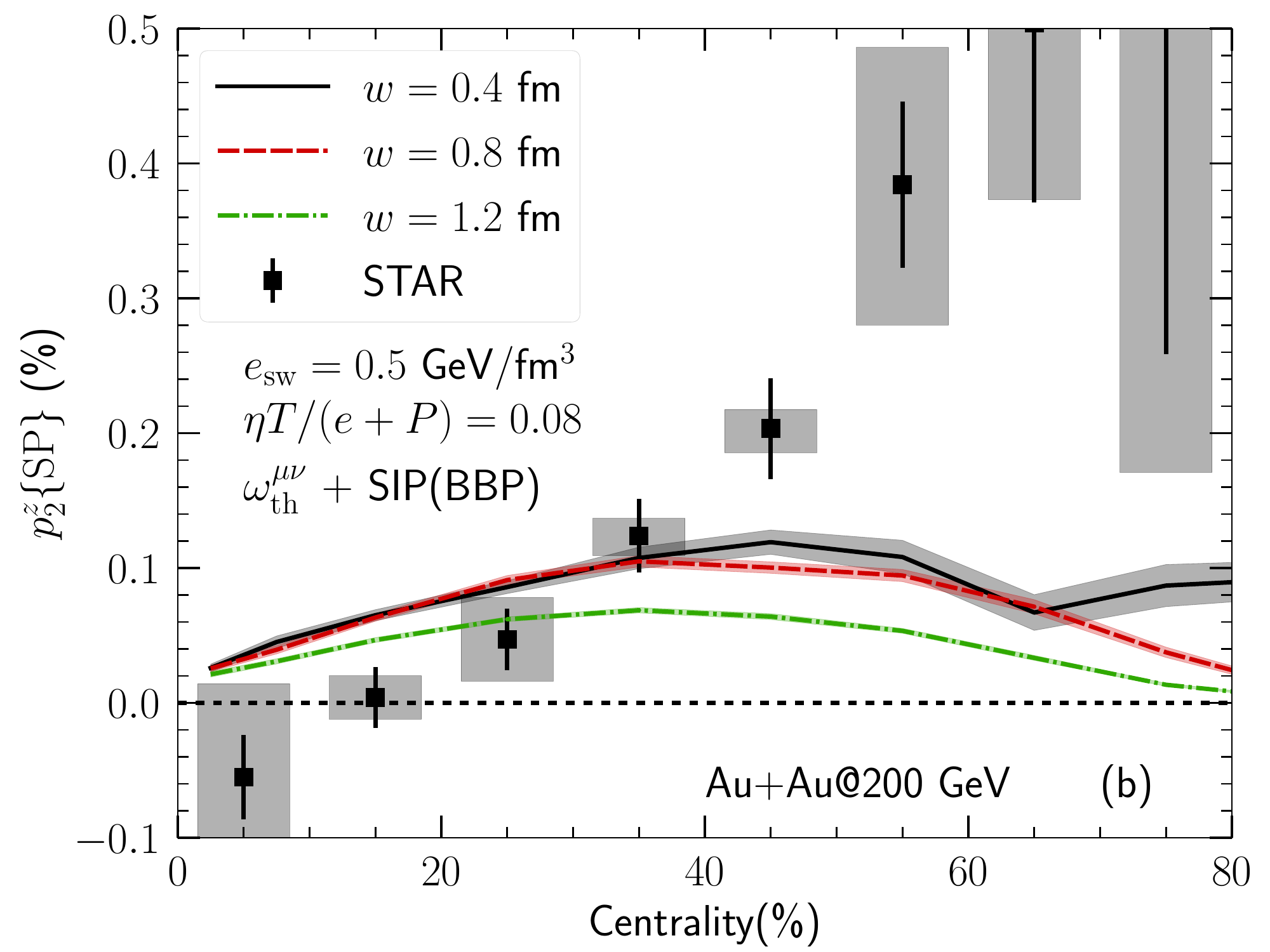}
    \includegraphics[width=0.9\linewidth]{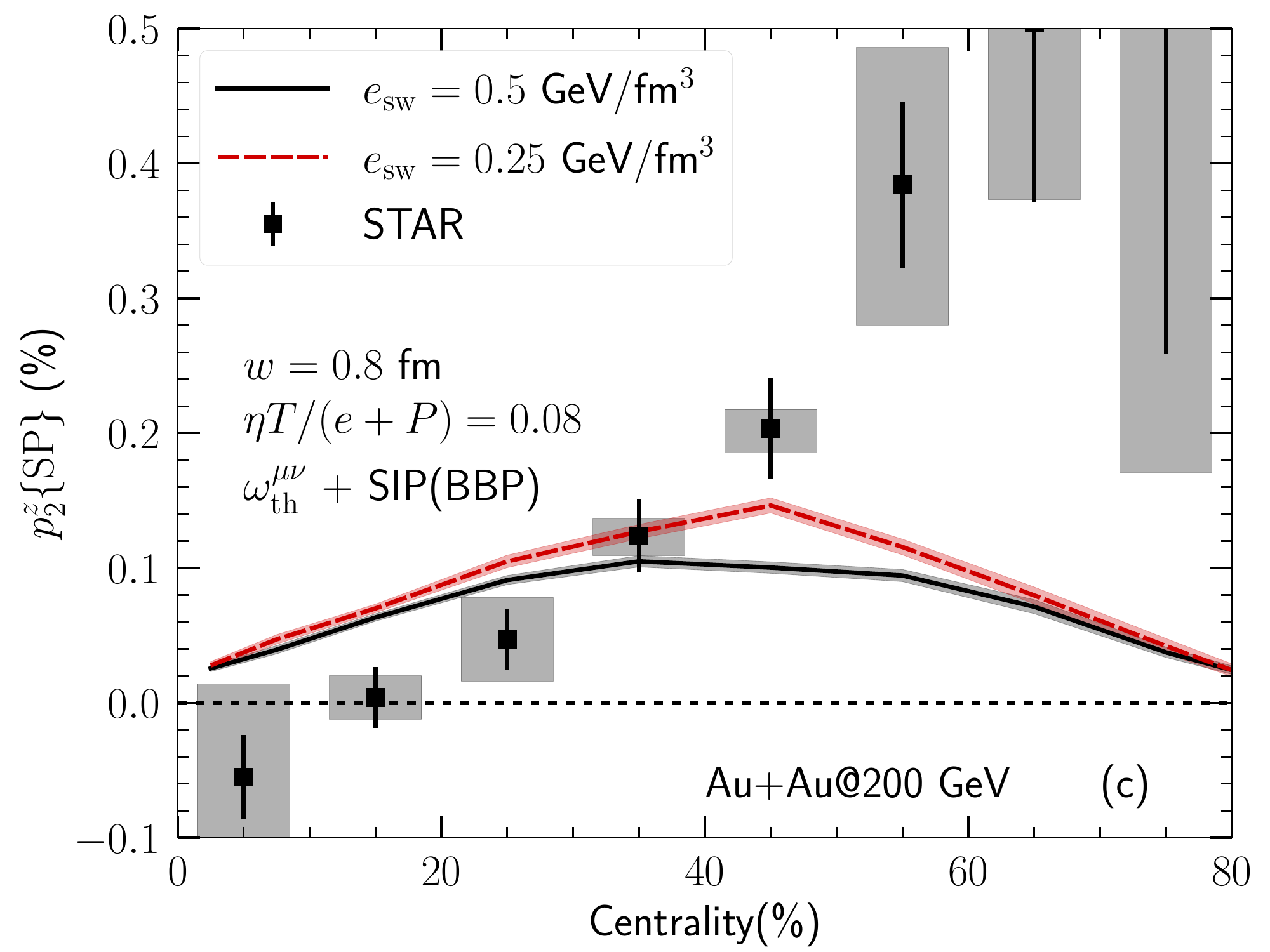}
    \caption{(Color Online) The centrality dependence of the second-order Fourier coefficients of the longitudinal polarization $P^z$ with respect to the elliptic flow event-plane angle $\Psi_2$ for different values of specific shear viscosity (a), initial hot spot size (b), and switching energy density (c). Results are compared with the STAR data \cite{STAR:2019erd}.}
    \label{fig:PzCenParamDep}
\end{figure}

In Figs.~\ref{fig:PzCenParamDep}, we systematically explore the sensitivity of the second-order Fourier coefficient of $P^z(\phi)$ on the QGP shear viscosity, initial hot spot size, and switching energy density. We find that the $p_2^z\{\mathrm{SP}\}$ increases with the value of specific shear viscosity used in the hydrodynamic phase. Ideal hydrodynamic simulations generate an almost zero $p_2^z\{\mathrm{SP}\}$, while the two finite values of shear viscosity give comparable $p_2^z\{\mathrm{SP}\}$ in central and semi-peripheral collisions. Figure~\ref{fig:PzCenParamDep}b shows that the $p_2^z\{\mathrm{SP}\}$ coefficient has a mild dependence on the initial hot spot size. Simulations with a large hot spot size $w = 1.2$ fm have a smaller $p_2^z\{\mathrm{SP}\}$ coefficient compare to those from simulations with the smaller $w$. Finally, Figure~\ref{fig:PzCenParamDep}c shows that a lower switching energy density $e_\mathrm{sw} = 0.25$\,GeV/fm$^3$ leads to a 15\% larger $p_2^z\{\mathrm{SP}\}$ compared to the results from simulations with $e_\mathrm{sw} = 0.5$\,GeV/fm$^3$. This result suggests that the coefficient $p_2^z\{\mathrm{SP}\}$ grows with the fireball lifetime. With all these combinations of model parameters, we find the values of $p_2^z\{\mathrm{SP}\}$ remain small in the peripheral Au+Au collisions beyond 50\% in centrality. It requires a more detailed analysis to resolve the difference with the experimental data in peripheral centrality bins. Compared to the sensitivity study for the $\Lambda$'s global polarization in Figs.~\ref{fig:PyParamDep}, the $p_2^z\{\mathrm{SP}\}$ coefficient of the longitudinal polarization does not show very strong sensitivity to the model parameters.

\begin{figure}[t!]
    \centering
    \includegraphics[width=0.9\linewidth]{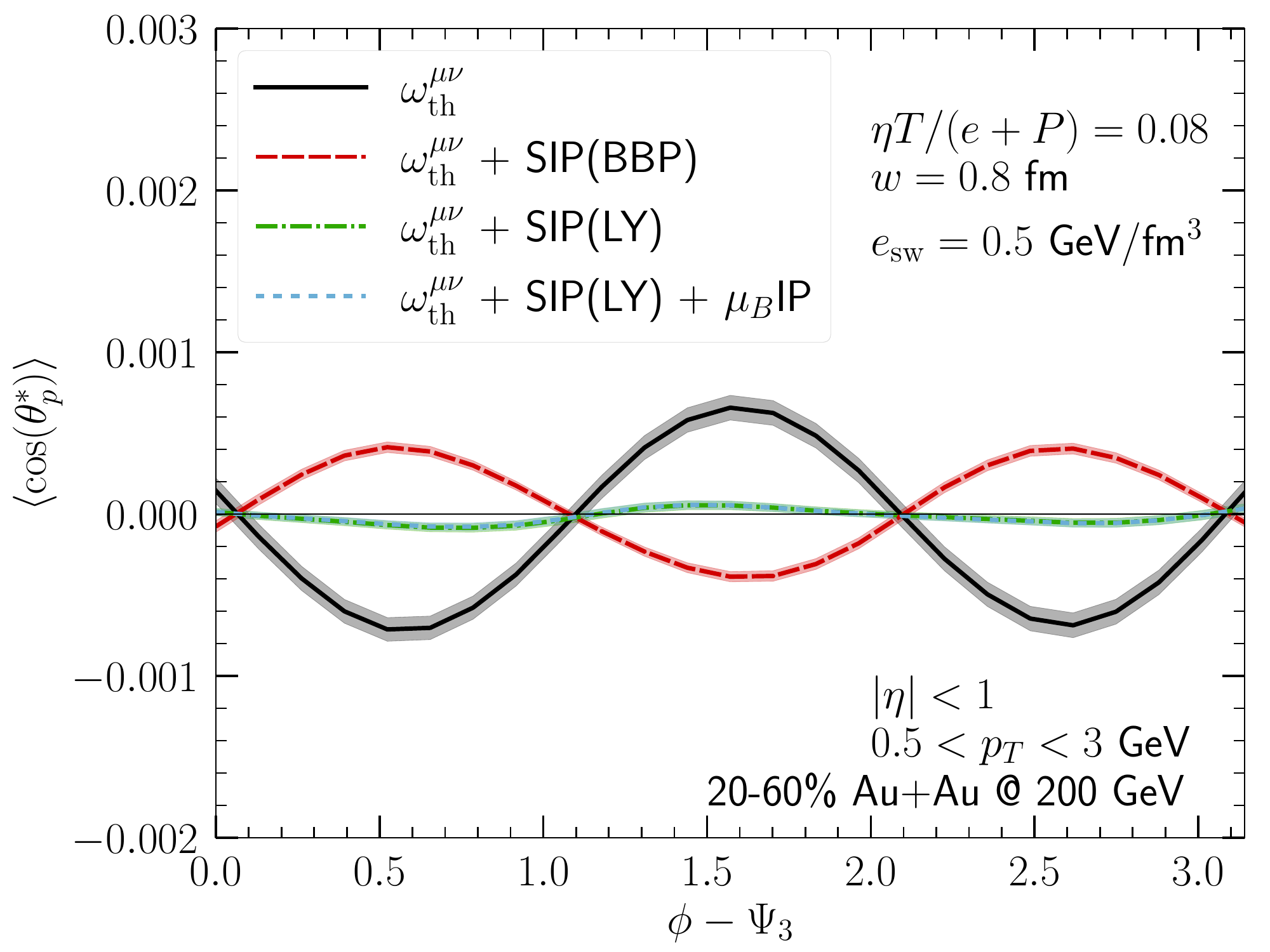}
    \caption{(Color Online) The $\langle \cos(\theta_p^*) \rangle$ in Eq.~\eqref{eq:costhetap} with respect to the third-order event plane angle computed from the $\Lambda$'s longitudinal polarization $P^z(\phi)$ using four combinations of the axial vector $\mathcal{A}^\mu$ in 20-60\% Au+Au collisions.}
    \label{fig:PzSIP_Psi3}
\end{figure}

Event-by-event simulations allow us to go beyond the second-order oscillation of the longitudinal polarization. We can compute higher-order Fourier coefficients of $P^z$ with respect to the event plane of higher-order anisotropic flow. Figure~\ref{fig:PzSIP_Psi3} shows an example of performing an event-average of the longitudinal polarization $P^z(\phi)$ with respect to the triangular flow event plane in 20-60\% Au+Au collisions. We can clearly see the third-order oscillation of the longitudinal polarization vector. Similar to the second-order case, the shear-induced polarization gives the opposite contributions to the azimuthal dependence compared to those from the thermal vorticity tensor. The SIP(BBP) term from Ref.~\cite{Becattini:2021suc} again gives a substantial contribution to flip the sign of $P^z$. Therefore, it is important to measure the third-order oscillation of the longitudinal polarization in experiments to further test whether this theoretical model is valid or not.

\begin{figure}[t!]
    \centering
    \includegraphics[width=0.9\linewidth]{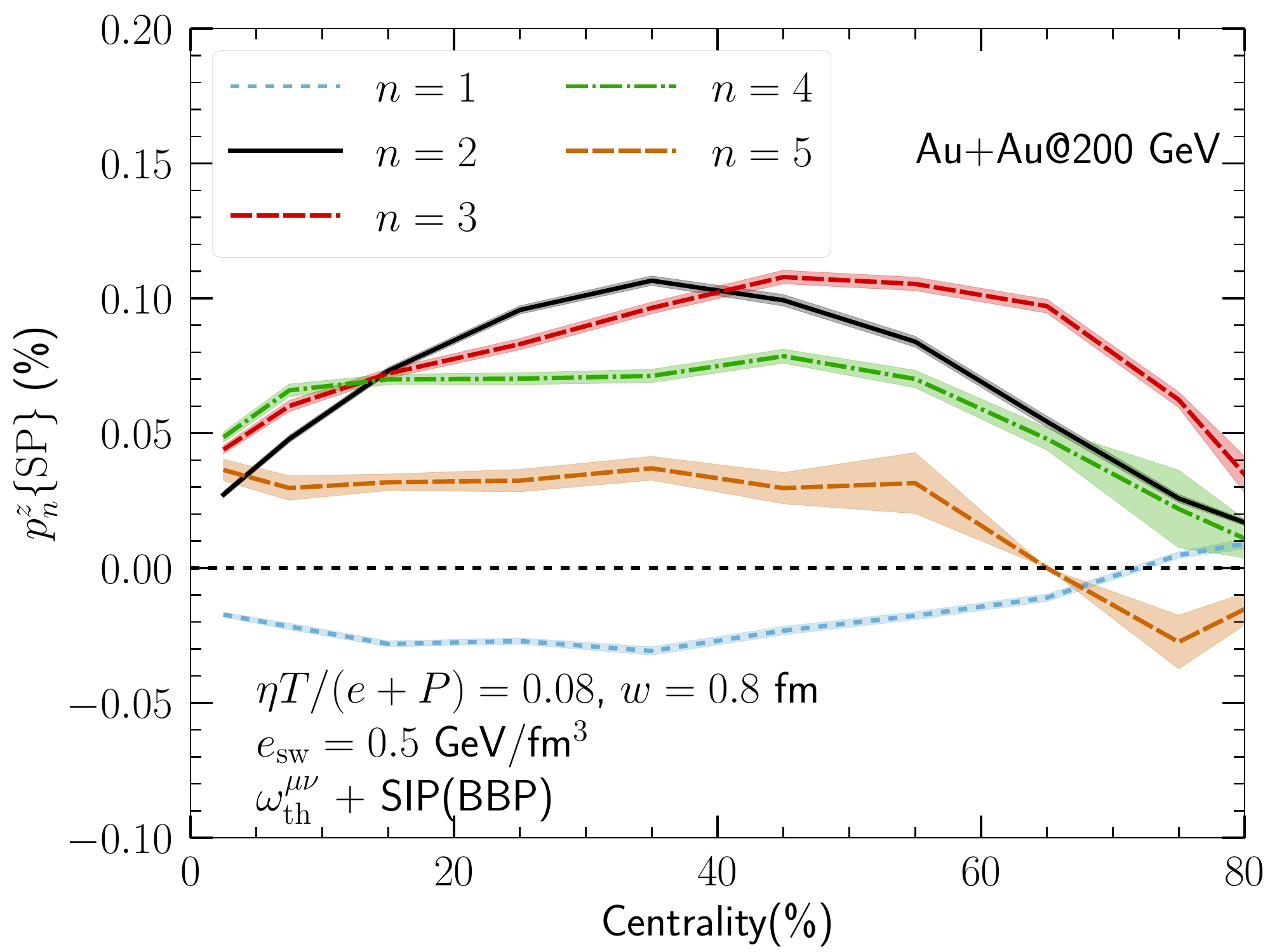}
    \caption{(Color Online) The centrality dependence of the $n$-th order Fourier coefficients of $P^z(\phi)$ with respect to $n$-th order event-plane determined by charged hadron anisotropic flow in Au+Au collisions at 200 GeV for $n = 1 - 5$.}
    \label{fig:Pzfn}
\end{figure}

In Fig.~\ref{fig:Pzfn}, we compute the scalar-product $p_n^z\{\mathrm{SP}\}$ between the Fourier coefficients of $P^z(\phi)$ and charged hadron anisotropic flow $v_n$ for $n = 1-5$ as functions of centrality in Au+Au collisions at 200 GeV. We find that the magnitudes of the third and the fourth order oscillations $p_3^z\{\mathrm{SP}\}$ and $p_4^z\{\mathrm{SP}\}$ are comparable to that of $p_2^z\{\mathrm{SP}\}$, while those of the $p_1^z\{\mathrm{SP}\}$ and $p_5^z\{\mathrm{SP}\}$ coefficients are small. The coefficient $p_1^z\{\mathrm{SP}\}$ computed with thermal vorticity + the SIP(BBP) contribution is negative for all centrality bins. We check that the shear induced polarization from Ref.~\cite{Becattini:2021suc} flips the signs of all orders of $p_n^z\{\mathrm{SP}\}$.
The centrality dependence of the $p_n^z\{\mathrm{SP}\}$ coefficients in Fig.~\ref{fig:Pzfn} provides a quantitative model prediction for the azimuthal dependence of longitudinal polarization and how it is correlated with the hydrodynamic anisotropic flow coefficients. Verifying these predictions in the experiments can help us further understand the origin of the $\Lambda$ spin polarization in heavy-ion collisions.

To further quantify the event-by-event correlation between the magnitudes of the anisotropic flow $v_n$ and the Fourier coefficients of the longitudinal polarization $P^z_n$, we can define the following
Pearson correlations,
\begin{equation}
    \rho(v_n^2, (P^z_n)^2) = \frac{\langle \hat{\delta}v_n^2 \hat{\delta}(P^z_n)^2 \rangle_\mathrm{ev}}{\sqrt{\langle (\hat{\delta}v_n^2)^2 \rangle_\mathrm{ev} \langle (\hat{\delta}(P^z_n)^2)^2 \rangle_\mathrm{ev}}},
    \label{eq:vnfnCorr}
\end{equation}
where $\langle \cdots \rangle_\mathrm{ev}$ represents the event average and the relative fluctuation of any observable $O$ is defined as,
\begin{equation}
    \hat{\delta} O = \delta O - \frac{\langle \delta O \delta N_\mathrm{ch} \rangle_\mathrm{ev}}{\langle (\delta N_\mathrm{ch})^2 \rangle_\mathrm{ev}} \delta N_\mathrm{ch} \quad \mbox{with} \quad \delta O = O - \langle O \rangle_\mathrm{ev}.
\end{equation}
Here the relative fluctuations subtract the correlation with the particle multiplicity in the event~\cite{Schenke:2020uqq}.

\begin{figure}[t!]
    \centering
    \includegraphics[width=0.9\linewidth]{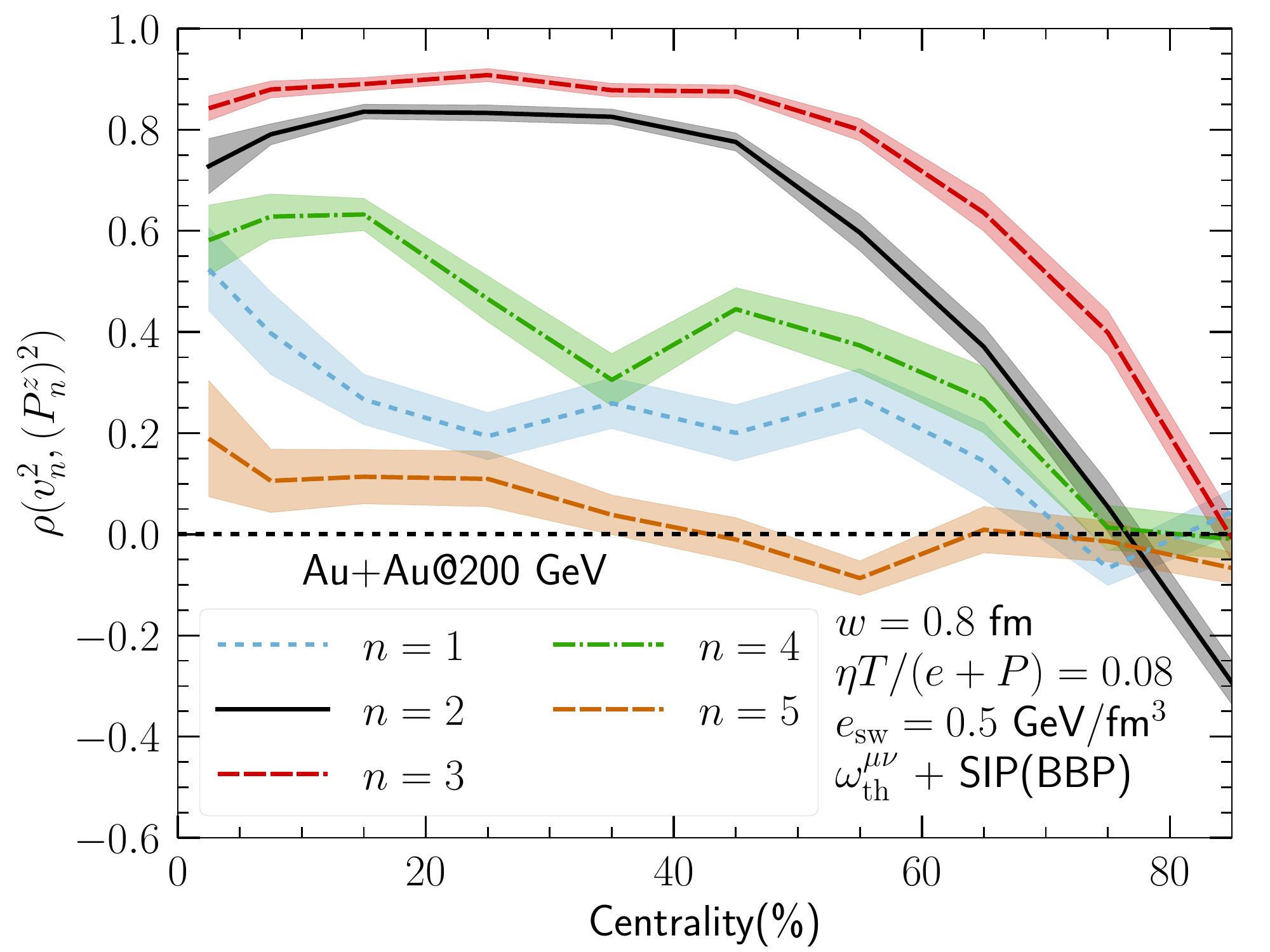}
    \caption{(Color Online) The centrality dependence of the Pearson correlation between charged hadron anisotropic flow $v_n$ and the Fourier coefficients of the longitudinal polarization $P^z_n$ in Au+Au collisions at 200 GeV.}
    \label{fig:vnfnCorr}
\end{figure}

In Fig.~\ref{fig:vnfnCorr}, we calculate the Pearson correlations between the magnitude of anisotropic flow $v_n$ and the $P^z_n$ coefficients for $n = 1-5$ in Au+Au collisions. We find strong positive correlations between $v_n$ and $P^z_n$ for $n = 2$ and $3$ from central up to 60\% in centrality. These strong correlations indicate that an event-shape analysis by selecting collision events according to $v_n$ at fixed multiplicity could show a positive correlation with the $P^z_n$ coefficient of the $\Lambda$'s longitudinal polarization.
The strengths of the $\rho(v_n^2, (P^z_n)^2)$ correlations are weaker for the harmonic orders $n = 1$ and $5$ compared to those for $n = 2-4$, but they are still significantly larger than zero.
Finally, these Pearson correlations are four-particle correlations that can be directly measured by the experiments provided there are enough statistics. These measurements can verify whether the event-by-event oscillation patterns of the longitudinal polarization are correlated with the underlying anisotropic flow of the medium.

\subsection{Collision system scan at the top RHIC energy}

The Relativistic Heavy-Ion Collider is a versatile machine to study the proprieties of Quark-Gluon Plasma with different nuclei species. In addition to Au+Au collisions, Ru+Ru and O+O collisions are recently performed with high precision. These smaller collision systems allow us to study how the hyperon's polarization observables change with the collision system size. In this subsection, we perform a parameter-free extrapolation from Au+Au collisions to these small systems. 

\begin{figure}[t!]
    \centering
    \includegraphics[width=0.9\linewidth]{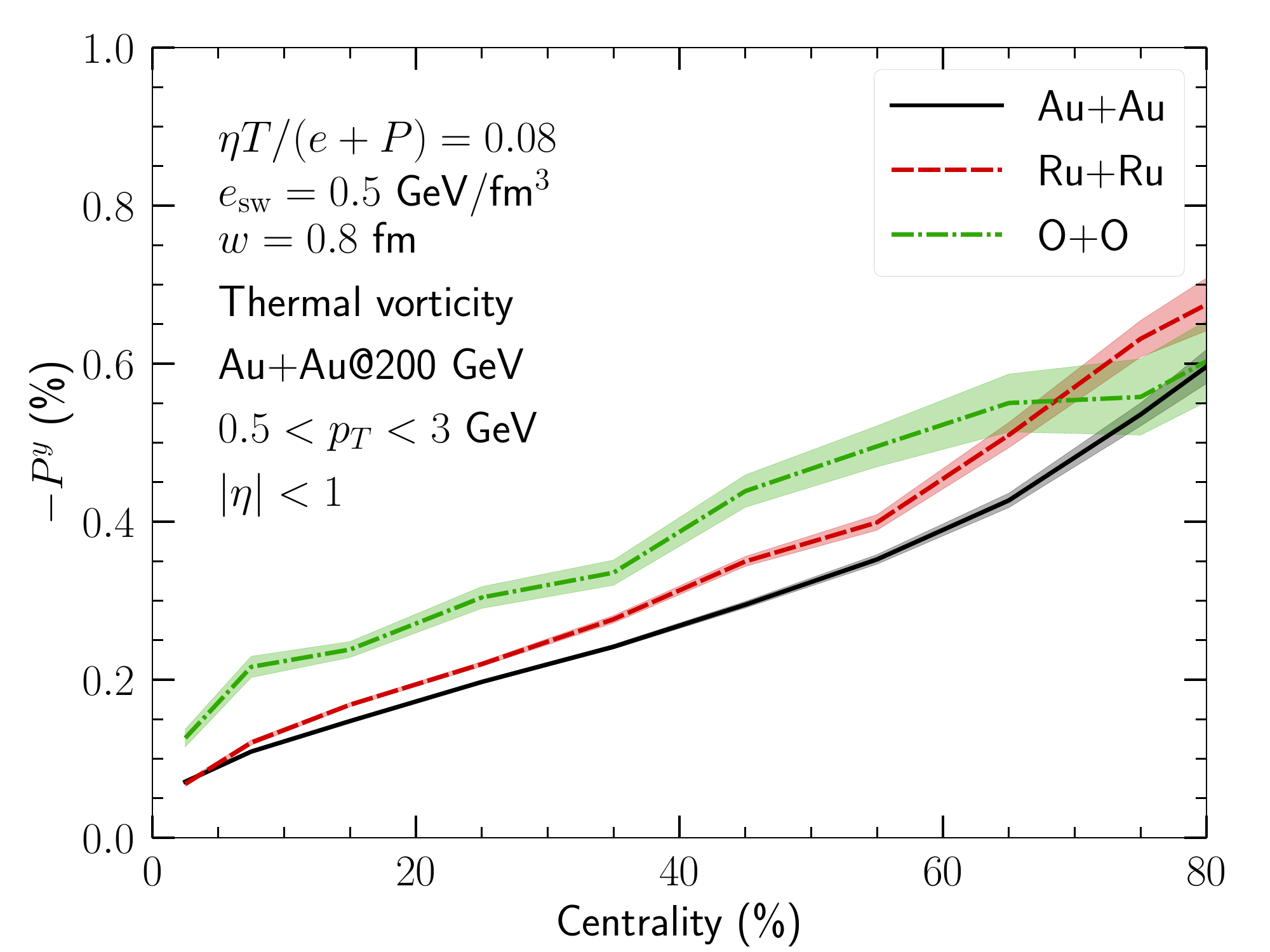}
    \caption{(Color Online) The centrality dependence of $\Lambda$'s global polarization in Au+Au, Ru+Ru, and O+O collisions at $\sqrt{s_\mathrm{NN}} = 200$\,GeV.}
    \label{fig:SysScanPy}
\end{figure}

Figure~\ref{fig:SysScanPy} shows the centrality dependence of the $\Lambda$ global polarization in Au+Au, Ru+Ru, and O+O collisions. For all three collision systems, the magnitude of the global polarization increases with the centrality. Because in our model the $\Lambda$ global polarization is strongly correlated with the system's orbital angular momentum, the monotonic increase of the $P^y$ with centrality reflects that the centrality defined by charged hadron multiplicity is strongly correlated with the collision geometry, even in the small O+O collisions. Now comparing the magnitudes of $P^y$ across the three collision systems, we find the $P^y$ is larger in smaller systems at the same centrality. This system size dependence can be understood as the fireball lifetimes are shorter in the smaller collision systems. Hence, thermal vorticity tensors have less time to reduce their sizes in O+O collisions than those in Au+Au collisions at the same centrality bin. With the future experimental measurements at RHIC, our prediction for the system size dependence of $\Lambda$'s $P^y$ can help us to verify whether the fluid thermal vorticity is the main contribution to $\Lambda$'s global polarization.

\begin{figure}[t!]
    \centering
    \includegraphics[width=0.9\linewidth]{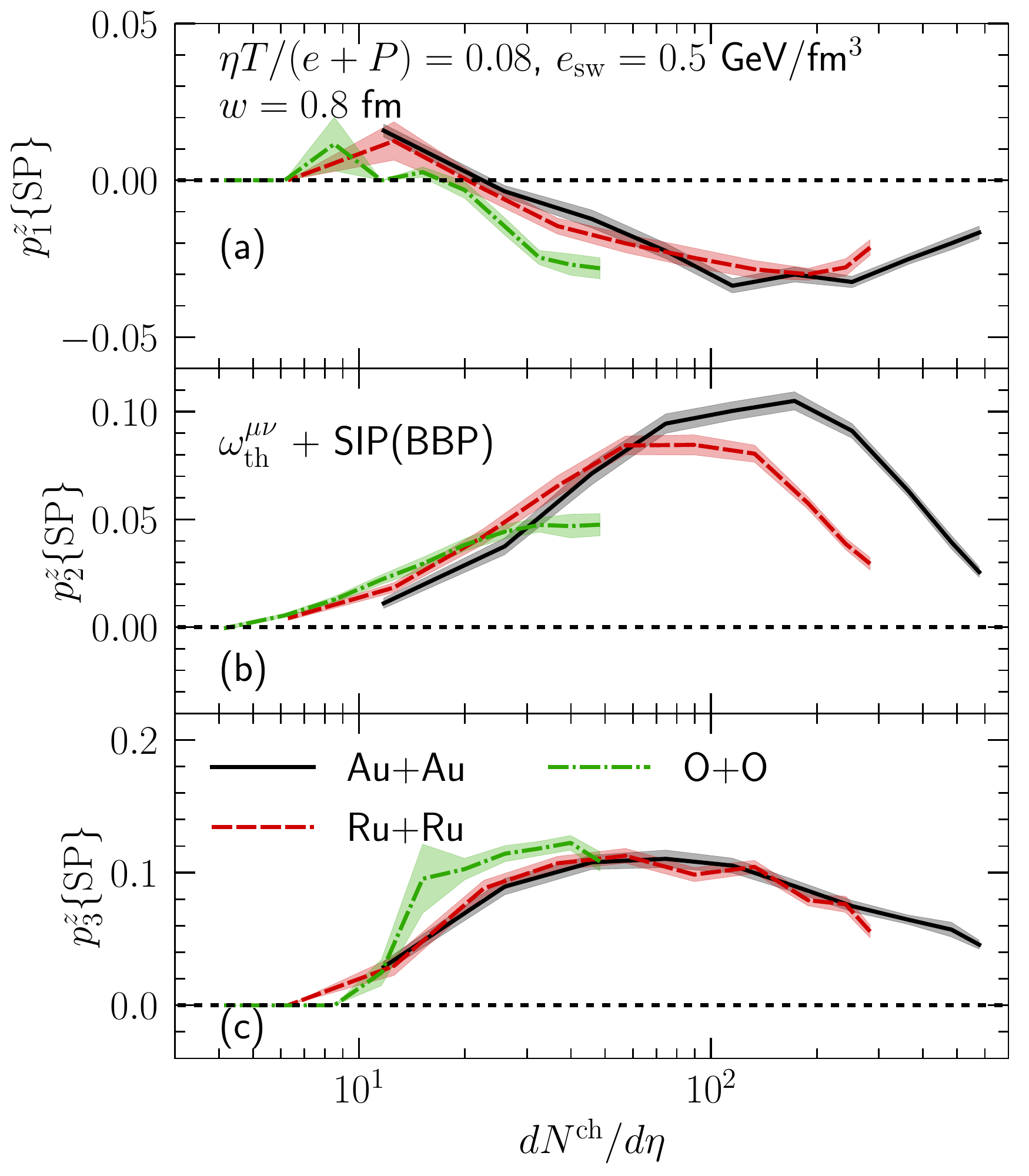}
    \caption{(Color Online) The Fourier coefficients of $\Lambda$'s longitudinal polarization as functions of charged hadron multiplicity for Au+Au, Ru+Ru, and O+O collisions at $\sqrt{s_\mathrm{NN}} = 200$\,GeV.}
    \label{fig:SysScanfn}
\end{figure}

We also study the system size dependence of the azimuthal dependence of the $\Lambda$ longitudinal polarization. In Fig.~\ref{fig:SysScanfn}, we plot the Fourier coefficients $p^z_n\{\mathrm{SP}\}$ of the $\Lambda$ longitudinal polarization for $(n = 1-3)$ in Au+Au, Ru+Ru, and O+O collisions. We have not accumulated enough statistics for the higher-order coefficients $n = 4$ and $5$ for the small collision systems.
We find a good multiplicity scaling for $p^z_1\{\mathrm{SP}\}$ and $p^z_3\{\mathrm{SP}\}$ across different collision systems. 
This multiplicity scaling is expected because both the dipolar and triangular flow coefficients are mostly driven by event-by-event fluctuations in all these three collision systems. The odd-order oscillations of $P^z$ are mainly controlled by the fireball lifetime which is correlated with the particle multiplicity. The multiplicity scaling does not work for $p_2^z\{\mathrm{SP}\}$ because it receives contributions from the collision geometry in central and semi-peripheral collisions.

\section{Conclusions} \label{sec:conc}

In this work, we perform event-by-event (3+1)D dynamical simulations to systematically study the hyperon's polarization observables in heavy-ion collisions at 200 GeV. 

We implement the two shear-induced polarization formulae proposed in the literature and quantify their effects on the $\Lambda$'s global and longitudinal polarization observables within the same (3+1)D dynamical framework. We find both shear-induced polarization terms give relatively small contributions to the $\Lambda$ global polarization near mid-rapidity. Therefore, the $\Lambda$'s global polarization is a robust observable to probe the thermal vorticity distribution in the dynamical system. 
In the forward rapidity region $\vert \eta \vert > 2$, the shear-induced polarization SIP(BBP) from Ref.~\cite{Becattini:2021suc} gives large contribution to $P^y(\eta)$.
Both shear-induced polarization terms give significant contributions to the azimuthal dependence of $\Lambda$'s global and longitudinal polarization. In particular, the SIP(BBP) from Ref.~\cite{Becattini:2021suc} can flip the sign of the longitudinal polarization $P^z(\phi)$. The main difference between the two SIP terms is that the SIP(BBP) term includes additional contributions from temperature gradients and fluid acceleration compared to the SIP(LY) term. Our model results with the SIP(BBP) contributions are different from those shown in Ref.~\cite{Becattini:2021iol}. The difference could come from the different types of initial conditions used in the simulations. In the meantime, our results with the SIP(LY) contributions show qualitative agreements with those in Ref.~\cite{Yi:2021unq, Fu:2021pok}.

With event-by-event simulations, we systematically study how to use the $\Lambda$ polarization observables to constrain the dynamical properties of relativistic heavy-ion collisions. The global polarization shows strong sensitivities to the initial-state fluctuations and the QGP's specific shear viscosity. It offers complementary information to anisotropic flow coefficients. Studying polarization and anisotropic flow observables together as functions of collision system size will set strong constraints on the 3D dynamics of heavy-ion collisions. With future more accurate measurements, the $\Lambda$ global polarization is an important observable to be included in the global Bayesian statistical analysis to constrain all aspects of the hot and dense nuclear matter.

The event-by-event simulations also provide us with a theoretical tool to make predictions for new correlation observables. We propose the four-particle Pearson correlation between charged hadron anisotropic flow $v_n$ and the Fourier coefficients of the longitudinal polarization $P_n^z$ for $n=1-5$. It will be exciting to verify the predicted correlations experimentally, which would open a new venue in the precision era of heavy-ion physics.

\acknowledgements
We thank Michael Lisa and Sergei Voloshin for the fruitful discussion.
This work is supported by the U.S. Department of Energy (DOE) under award numbers DE-SC0021969 and DE-SC0013460.
CS acknowledges a DOE Office of Science Early Career Award. SA acknowledges scholarship supports from the Department of Physics, Jazan University, Jazan, Kingdom of Saudi Arabia.
This work is in part supported by the U.S. Department of Energy, Office of Science, Office of Nuclear Physics, within the framework of the Beam Energy Scan Theory (BEST) Topical Collaboration.
This research was done using resources provided by the Open Science Grid (OSG) \cite{Pordes:2007zzb, Sfiligoi:2009cct}, which is supported by the National Science Foundation award \#2030508.

\bibliography{AAebeVorticity, nonInspire}

\end{document}